\documentclass[preprint,12pt]{aastex}  

\def\cm{\ifmmode {\rm cm}^{-1} \else cm$^{-1}$ \fi}
\def\s{\ifmmode {\rm s}^{-1} \else s$^{-1}$ \fi}
\def\cc{\ifmmode {\rm cm}^{-3} \else cm$^{-3}$ \fi}
\def\cs{\ifmmode {\rm cm}^{-2} \else cm$^{-2}$ \fi}
\def\g{\ifmmode \gamma \else $\gamma$\fi}
\def\G{\ifmmode \Gamma \else $\Gamma$\fi}
\def\Gs{\ifmmode \Gamma~ \else $\Gamma~$\fi}

\def\gc{\ifmmode \gamma_{\rm c} \else $\gamma_{\rm c}$ \fi}
\def\sw{Schwarzschild~}
\def\gsim{\mathrel{\raise.5ex\hbox{$>$}\mkern-14mu
             \lower0.6ex\hbox{$\sim$}}}
\def\lsim{\mathrel{\raise.3ex\hbox{$<$}\mkern-14mu
             \lower0.6ex\hbox{$\sim$}}}
\def\simless{\mathbin{\lower 3pt\hbox
     {$\rlap{\raise 5pt\hbox{$\char'074$}}\mathchar"7218$}}}   
\def\simmore{\mathbin{\lower 3pt\hbox
     {$\rlap{\raise 5pt\hbox{$\char'076$}}\mathchar"7218$}}}   
\def\Msun{M_\odot}                                
\def\4u{4U 1728--34}
\def\deg{^\circ}

\received{} \accepted{}
\slugcomment{Accepted to ApJ, Feb. 5, 2008: v4}

\lefthead{et al.} \righthead{\4u}

\shorttitle{HFQPOs from Random X-ray Bursts}

\shortauthors{Fukumura \& Kazanas}

\begin{document}

\title{Light Echoes in Kerr Geometry: A Source of High Frequency QPOs from Random
X-ray Bursts}


%
\author{Keigo Fukumura \& Demosthenes Kazanas}
\affil{Astrophysics Science Division, NASA Goddard Space Flight
Center, Code 663, Greenbelt, MD 20771}
\email{Keigo.Fukumura@nasa.gov, Demos.Kazanas@nasa.gov}
%
%

\begin{abstract}

\baselineskip=15pt

We propose that high frequency quasi-periodic oscillations (HFQPOs)
can be produced from randomly-formed X-ray bursts (flashes) by
plasma interior to the ergosphere of a rapidly-rotating black hole.
We show by direct computation of their orbits that the photons
comprising the observed X-ray light curves, if due to a multitude of
such flashes, are affected significantly by the black hole's
dragging of inertial frames; the photons of each such burst arrive
to an observer at infinity in multiple (double or triple), distinct
``bunches" separated by a roughly constant time lag of $\Delta
t_{\rm lag}/M \simeq 14$, regardless of the bursts' azimuthal
position. We argue that every other such ``bunch" represents photons
that follow trajectories with an additional orbit around the black
hole at the photon circular orbit radius (a photon ``echo"). The
presence of this constant lag in the response function of the system
leads to a QPO feature in its power density spectra, even though the
corresponding light curve consists of a totally stochastic signal.
This effect is by and large due to the black hole spin and is shown
to gradually diminish as the spin parameter $a$ decreases or the
radial position of the burst moves outside the static limit surface
(ergosphere). Our calculations indicate that for a black hole with
Kerr parameter of $a/M=0.99$ and mass of $M=10\Msun$ the QPO is
expected at a frequency of $\nu_{\rm QPO} \sim 1.3-1.4$ kHz. We
discuss the plausibility and observational implications of our
model/results as well as its limitations.

\end{abstract}

\keywords{accretion, accretion disks --- black hole physics ---
X-rays: galaxies --- stars: oscillations  }

\baselineskip=15pt

\section{Introduction}


Following the inspiring observational discoveries of quasi-periodic
oscillations (QPOs) from compact objects [see, e.g., van der Klis
2000 for neutron star low-mass X-ray binaries; Strohmayer 2001a,b
and Cui et al.~1999 for stellar-mass black hole systems; Strohmayer
et al.~2007 for ultraluminous X-ray sources (ULXs)], a number of
theoretical scenarios have been proposed to explain the physics
behind the QPO observations. These models are based, among others,
also on the dynamics of relativistic accretion disks, especially for
the QPOs of  systems thought to harbor black holes. For example, it
has been proposed that the X-ray modulation giving rise to the
observed QPOs can be produced at the precession frequency of
accretion disks due to relativistic dragging of inertial frames
around rapidly-rotating black holes \citep[][]{LT18}, with the QPO
frequency used to estimate the range of the hole's spin parameter
\citep[e.g.][for XTE~J1550-564, GRO~J1655-40,~GRS~1915+105, Cyg X-1,
and GS~1124-68]{Cui98,Aschenbach04,Schnittman04,Schnittman06}. On
the other hand, resonance frequencies among various diskoseismic
oscillation modes \citep[e.g.][]{Nowak97,Kato01} have been invoked
to explain the observed 2:3 frequency commensurability in
XTE~J1550-564 and GRO~J1655-40
\citep[e.g.][]{Abramowicz01,Kluzniak01,Schnittman04,Donmez07}. More
recently, \citet{Aoki04} argued that radially propagating shocks in
relativistic accretion disks could produce the observed QPOs from
black hole systems (GRS~1915+105 and GRO~J1655-40). Finally,
considerations of inhomogeneities of accretion disks (e.g., due to
local magnetic flares and/or orbiting clumps) have led some authors
to investigate orbiting hot spot models \cite[e.g.][]{Karas99}. The
models in the above (necessarily incomplete) list can provide
frequencies in agreement with those of the observed QPOs by the
judicious choice of some of the systems' dynamical parameters. The
QPO features in these models rely on an underlying oscillatory
behavior of the light curve which, however, can be easily obscured
in the presence of noise, even though the QPO is clearly present in
the power spectra.

In this paper, we discuss a process for QPO formation that relies
not on an underlying (but noise-obscured) oscillation, but on an
``echo" of the input signal (the QPO behavior then follows from a
well known theorem of Fourier analysis). As such, we show that QPO
features are possible even for a totally random signal, which in
itself differs little from white noise. We show that such an ``echo"
(and the concomitant QPO) is possible for photons emitted by an
accretion disk whose Innermost Stable Circular Orbit (ISCO) reaches
within the ergosphere of a rotating black hole, even if the photon
emission is random in time, orbital phase and isotropic in the
plasma frame. Our analysis of this process indicates that these
features imply the presence of, unobserved as yet, high frequency
QPOs (HFQPOs), which rely crucially on the dragging of inertial
frames and are absent for slowly rotating black holes.

Our paper is structured as follows: In \S 2 we provide a general
description of the model, details of the photon kinematics and the
response function of the system. In \S 3 we give a prescription for
constructing stochastic model light curves and show that their power
density spectra (PDS) exhibit the QPO features as anticipated.
Finally, in \S 4 we review our results, make contact with
observations and discuss prospects of future work. We show the
derivation of the radial null geodesic equation in the Appendix.

\section{Description of the Model}

The model we consider consists of the standard geometrically thin,
optically thick accretion disk \citep[e.g.][]{Novikov73,Page74} that
extends to the ISCO. We assume the production of instantaneous
X-rays {\sl in short bursts} (short compared to the local orbital
period) at small heights above the disk as a result of either flares
from the reconnection of magnetic field loops anchored in the disk
\citep[e.g.][]{Galeev79,Haardt94,Poutanen99, Nayakshin01, Czerny04},
or of standing shocks \citep[e.g.][]{Fukumura07}. To simplify our
calculations we consider observers at small latitudes and we can
thus restrict with good accuracy the computation of photon orbits on
the equatorial plane ($\theta=\pi/2$). We assume that the X-ray
flares occur along the circumference of a radially thin annulus
randomly both in position and in time.
The small height of the disk and the source of the X-ray flares
guarantee that except for the photons intercepted by the disk, the
remainder can reach unimpeded the observer at infinity at nearly
equatorial orbits. We assume that the photons are emitted
isotropically in the rotating plasma rest frame [this entails some
subtleties \citep[see e.g.][]{FukKaz07}] and their trajectories are
influenced by both the motion of the emitter and the dragging of
inertial frames (the marginal orbit is inside the ergosphere for
sufficiently large values of the black hole spin).
We collect the photons at a large radial distance ($r_{\infty}/M =
600$ where $M$ is black hole mass) as a function of time for
different relative positions between the X-ray flare source and the
observer to compile the response function of the system for
observers at small disk latitudes.

To compute the photon orbits we adopt the Kerr metric in
Boyer-Lindquist (BL) coordinates ($r,\theta,\phi$) and geometrized
units ($G=c=1$ where $G$ is the gravitational constant and $c$ is
the speed of light). Distance and time are normalized by black hole
mass $M$ [i.e. $r  = 1.5 \times 10^5 \, (M/M_{\odot})$ cm and $t =
5.0 \times 10^{-6} \, (M/M_{\odot})$ s]. The accretion disk extends
from an unspecified outer radius to an inner radius (of ISCO) at
$r=r_{\rm ms}$, within which the matter freely spirals in toward the
event horizon. Hence, we assume that the X-ray source at
($r_s,\phi_s$) in the disk region follows the Keplerian motion,
while the source inside the ISCO must have radial and azimuthal
motion with its energy and angular momentum at the ISCO being
conserved subsequently. Plunging motion measured in a locally
non-rotating reference frame (LNRF) is described by
\begin{eqnarray}
v^\phi &=& \frac{A_s}{\Sigma_s \Delta_s^{1/2}} (\Omega_s-\omega) \ ,
\\
v^r &=& \frac{A_s^{1/2}}{\Delta_s} \frac{u^r}{u^t} \ ,
\label{eq:free-fall}
\end{eqnarray}
where $\Delta_s \equiv r_s^2-2Mr_s+a^2$, $A_s \equiv
(r_s^2+a^2)^2-a^2 \Delta_s^2$ and $a$ denotes a dimensionless Kerr
parameter. The angular velocity of the source is given by $\Omega_s
\equiv u^\phi/u^t$ while $\omega$ describes the angular velocity due
to the dragging of inertial frames, and the four-velocity
components, $u^t,~u^r$ and $u^\phi$, satisfy the normalization
condition ($u_\mu u^\mu =-1$). For Keplerian motion we have
$\Omega_s = \Omega_K = M^{1/2}/(a M^{1/2}+r_s^{3/2})$ and $u^r=0$.
We numerically solve the exact null trajectories by specifying the
position of a burst $(r_s,\phi_s)$ and the photon's impact parameter
$b$ (or specific axial angular momentum), employing the following
equations of motion \citep[see, e.g., Appendix in][]{Chandra83}
\begin{eqnarray}
\dot{t} &\equiv& \frac{d t}{d \lambda} = \frac{1}{\Delta} \left(r^2
+a^2 +\frac{2 a^2 M}{r} -\frac{2aM b}{r} \right) \ , \label{eq:tdot}
\\
\dot{\phi} &\equiv& \frac{d \phi}{d \lambda} = \frac{1}{\Delta}
\left[\frac{2aM}{r}+ \left(1-\frac{2M}{r}\right) b \right] \ ,
\label{eq:phidot}
\\
\dot{r} &\equiv& \frac{dr}{d \lambda} = \frac{\xi^{1/2} \cos^2 \delta
\left(\pm r \Delta |\sec \delta| - 2 a M \Delta^{1/2} \tan \delta
\right)} {r^{1/2} \left[r^3(r-2M)+a^2(r^2-2M^2)+2a^2M^2
\cos(2\delta)\right]} \ , \label{eq:rdot}
%
%
\end{eqnarray}
where $\Delta \equiv r^2 -2Mr +a^2$, $\xi \equiv r^3+a^2(r+2M)$,
$\lambda$ is the null affine parameter, and $\delta$ measures
photon's emission angle between propagation direction and radial
direction \citep[e.g.,][p. 675]{MTW73} in the LNRF (as opposed to the
rotating fluid frame).
%
%
See Appendix for the derivation of radial component of null
geodesics \citep[also][for a similar discussion of poloidal
geodesics]{FukKaz07}. We must take into account the azimuthal
beaming effects (photon focusing) on the emitting angle $\delta$ at
the source position ($r=r_s$), particularly in the innermost regions
($r \gtrsim r_{\rm ms}$) where the Doppler effects become
significant. This is defined as
\begin{eqnarray}
\delta = \arctan \left\{\frac{\sin \alpha}{\gamma
(v+\cos \alpha)}\right\} \ , \label{eq:delta}
\end{eqnarray}
where $\alpha$ measures photon's emission angle between propagation
direction and radial direction in the rest-frame of the source (as
opposed to LNRF), $v$ denotes the total three-velocity of the source
measured in LNRF, and $\gamma \equiv 1/\sqrt{1-v^2}$. We stress that
both quantities, $\alpha$ and $\delta$, are defined as photon's
angles between emission vector and radial component vector in two
different frames, which are related through the Doppler boost by
equation~(\ref{eq:delta}). In our computations $\alpha$ (rather than
$\delta$) is chosen uniformly in fluid frame for isotropic photon
emission. In the Appendix we explicitly show the definition of
$\delta$ and its relation to the photon's impact parameter $b$. Note
that $\delta \rightarrow \alpha$ as $v \rightarrow 0$ (i.e. no
beaming case) as expected. We then consider locally isotropic
emission from each burst in the fluid frame and trace the individual
photon ray by solving the geodesic equations in Kerr geometry.

\begin{figure}[t]
\epsscale{0.99} \plottwo{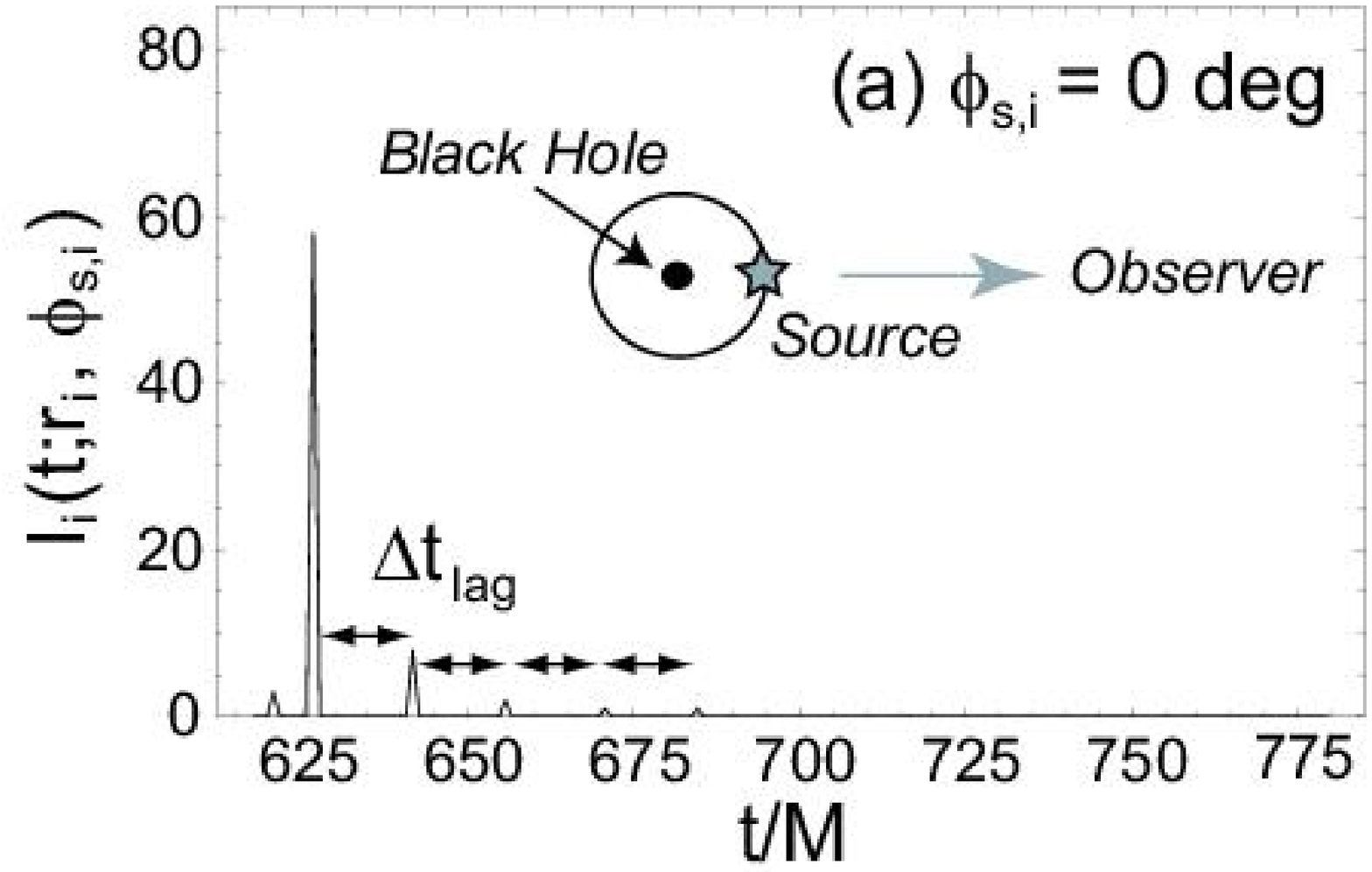}{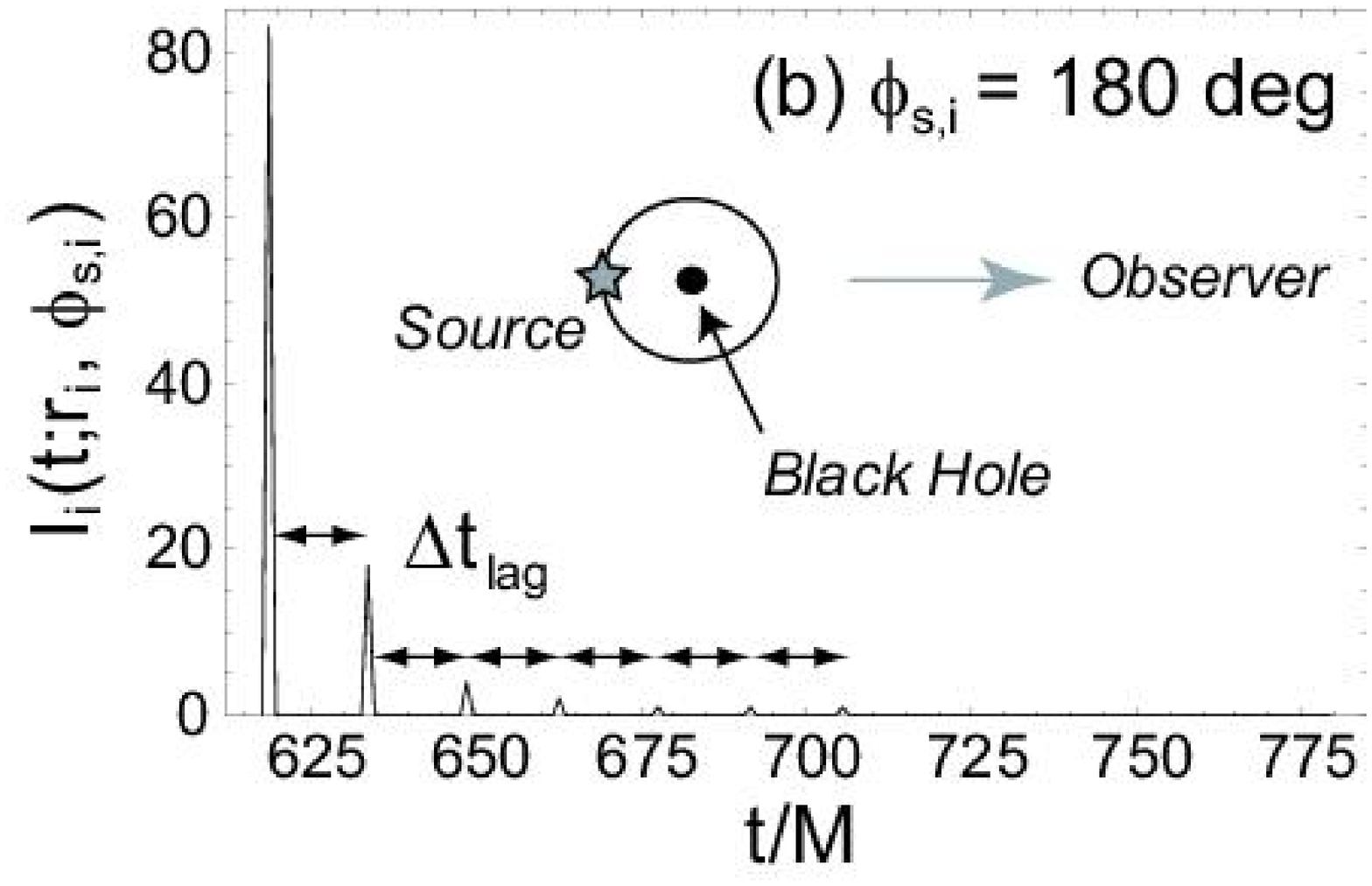} \caption{Sample signals
from two random X-ray bursts (out of a total of $N_b=6000$) at (a)
$\phi_{s,i} = 0\deg$ and (b) $\phi_{s,i}=180\deg$ for $a/M=0.99$ and
$r_s=r_{\rm ms}$. In each phase $N_{\rm ph}=200,000$ photons are
isotropically emitted. Arrows indicate a roughly constant time lag
$\Delta t_{\rm lag}/M \simeq 14$. Relative positions of the
observer, source and the black hole are illustrated.
\label{fig:response-a099}}
\end{figure}

\begin{figure}[htb]
\epsscale{0.4} \plotone{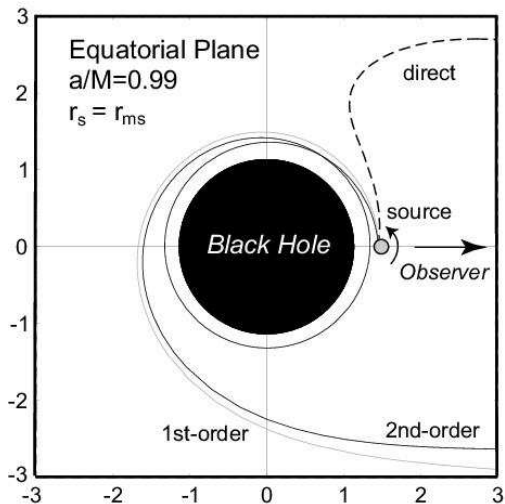}\caption{Representative photon rays
corresponding to the response function in
Figure~\ref{fig:response-a099}a. The trajectory in the upper right
quadrant (dashed curve) is a backward emitted photon. The gray and
dark curves (solid curves) represent photons that reach the observer
after $2\pi$ and $4 \pi$ radians, respectively. \label{fig:ray}}
\end{figure}

For a given source position $(r_s,\phi_{s,i})$ (where $\phi_{s,i} =
0$ for the source located between the observer and the black hole)
we collect the times $t_j$ of the photons arriving at $r =
r_{\infty} = 600 M$ within an angular bin of $\Delta \phi =
1^{\circ}$ centered at the observer, to produce the response
function of the system for the specific source location
$I_i(t;r_s,\phi_{s,i})$ [photon/time]. A total of $N_{\rm
ph}=200,000$ photon rays were computed from a single source. By
varying the phase $\phi_{s,i}$ of the source relative to the
observer between 0 and $2 \pi$ we then compiled the response
function of the system for all $\phi_{s,i}$ for a given $r_s$, and
the obtained photons were binned in time bins of size $M$. Samples
of this response for (a) $\phi_{s,i} = 0^{\circ}$ and (b)
$180^{\circ}$ are given in Figure \ref{fig:response-a099} for a
black hole with $a/M=0.99$ and emission from the marginally stable
orbit, i.e. $r_s = r_{\rm ms} \simeq 1. 45 \, M$. One can plainly
see that the response function consists of a series of pulses
separated by approximately $\Delta t_{\rm lag}/M \simeq 14$,
regardless of the azimuthal position of the source. These represent
photons arriving at the observer either directly or correspondingly
after one or more orbits around the black hole. Also, although the
observer is physically closer to the source in (a) than in (b),
he/she detects the signals emitted in (b) earlier. This is due to
the combination of the strong Doppler beaming due to the rotation of
the source and the dragging of inertial frames that send most
photons in case (a) around the black hole rather than directly into
the observer's line of sight. Sample of photon trajectories emitted
from a source with $\phi_{s,i}=0^\circ$ (corresponding to
Fig.~\ref{fig:response-a099}a) reaching the observer are shown in
Figure~\ref{fig:ray}.

In fact, Figure~\ref{fig:ray} encapsulates the essence of the effect
discussed herein: Because of the frame-dragging effects for a source
within the ergosphere, even those photons with negative angular
momenta (going backward in the local frame with respect to the
hole's rotation) are forced to propagate in the same sense as the
rotating hole (the photon escaping directly to the observer in the
upper right quadrant of Fig.~\ref{fig:ray} is one of them). All
other photons can reach the observer (at phase $\phi_{s,i} =
0^\circ$ in Fig.~\ref{fig:ray}) only by moving in the direction of
the hole's rotation; as such they can do so by moving by an angle
that is a fraction $q$ of $2 \pi$ ($q = 0$ for dashed ray; $q=1$ for
gray ray, and $q=2$ for solid dark ray) to produce the response of
Figure~\ref{fig:response-a099}a.
%
%
As we show explicitly in the next section, it is the constancy of
this time-lag that is responsible for the presence of QPOs in the
system. Such a constant lag is almost absent, however, for photon
sources outside the ergosphere or in the \sw geometry, leading to
qualitatively different results in these cases.
%

\begin{figure}[t]
\epsscale{0.99} \plottwo{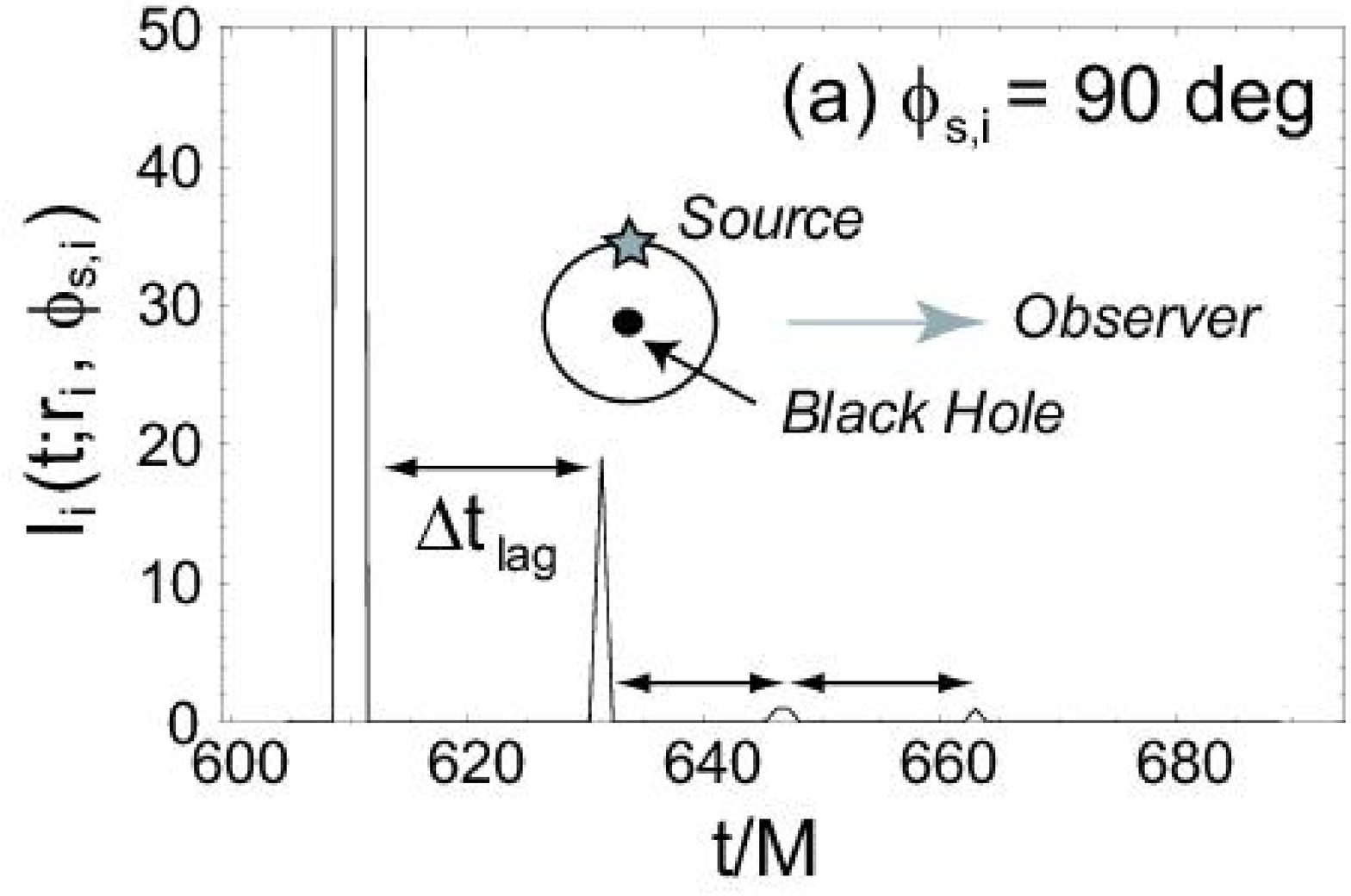}{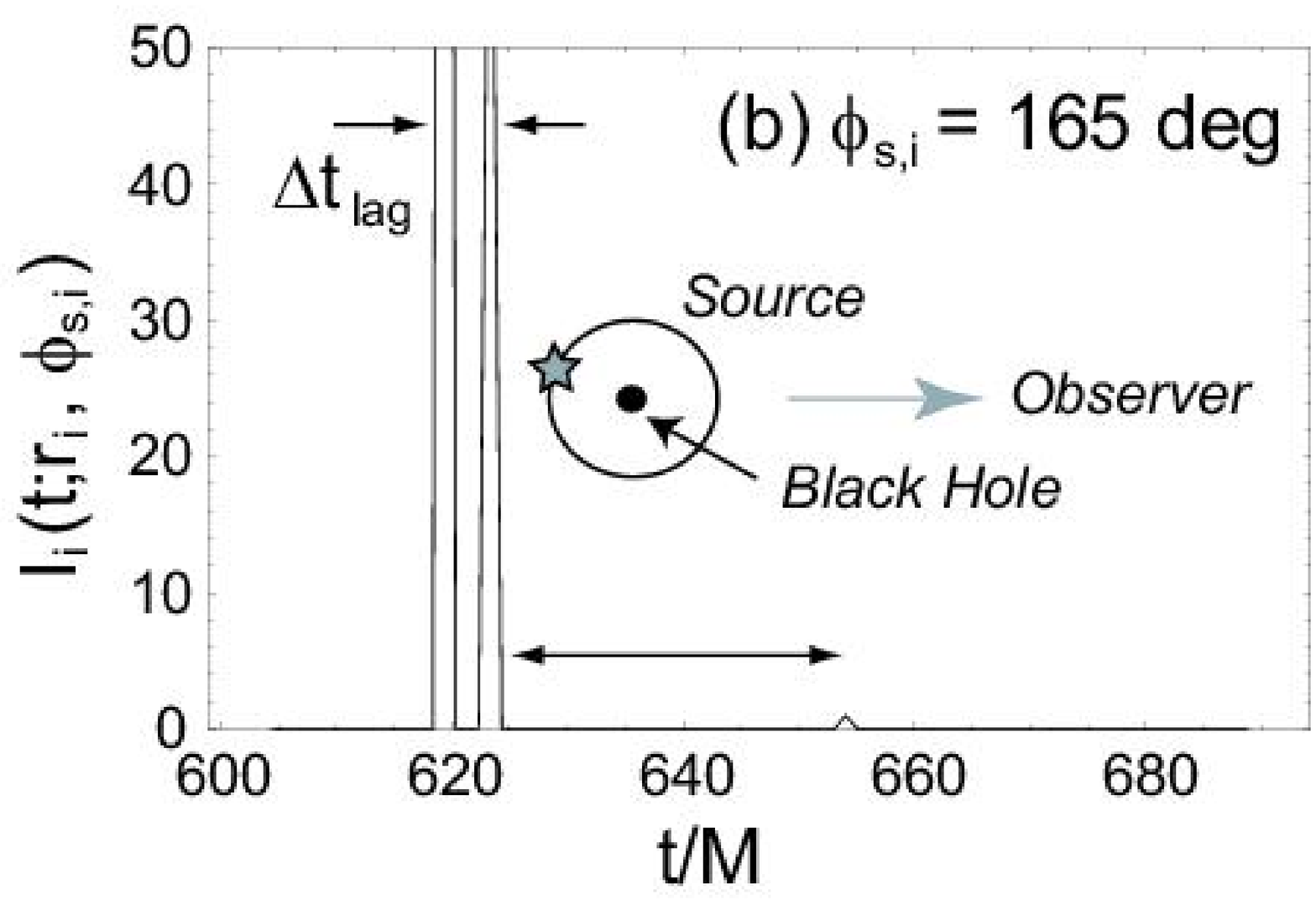} \caption{Same as
Figure~\ref{fig:response-a099} but for $a/M=0$ and $r_s=r_{\rm
ms}=6M$. No constant time-lag is seen. \label{fig:response-a0}}
\end{figure}

\begin{figure}[ht]
\epsscale{0.4} \plotone{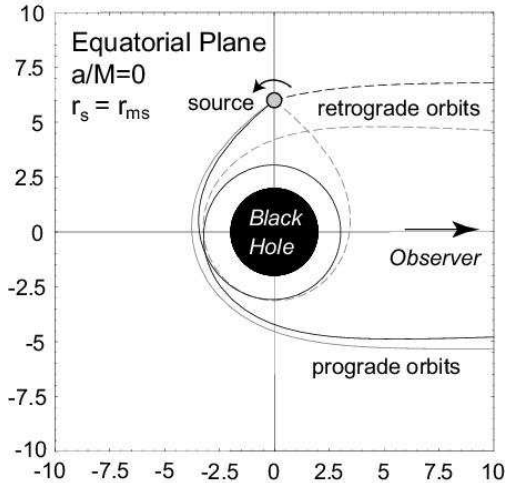} \caption{Representative photon rays
corresponding to the response function in
Figure~\ref{fig:response-a0}a. The trajectories in the upper right
quadrant (dashed curves) are backward emitted photon. The gray and
dark curves (solid curves) represent photons that reach the observer
after $\pi+\pi/2$ and $3\pi+\pi/2$ radians, respectively.
\label{fig:ray-a0}}
\end{figure}

To be sure, even in the \sw geometry one expects that some photons can
reach the observer after going around the hole by an angle that is
an integer-multiple of $2\pi$, but their number is much too small to
make an observable contribution to the response function.
In support of this claim we present similar computations in a \sw
geometry in which the frame-dragging effect is absent and the
marginally stable orbit is larger ($r_{\rm ms}/M=6$). The response
functions for (a) $\phi_{s,i}=90^\circ$ and (b) $165^\circ$ are
depicted in Figure~\ref{fig:response-a0} for $N_{\rm ph}=200,000$
photons. It is apparent that the response function in this case
comprises a very large peak at time $t/M \simeq 610$, a much smaller
one at $t/M \simeq 632$ and two additional even smaller ones
separated from this last one by $\Delta t_{\rm lag}/M \simeq 16$ and
$\Delta t_{\rm lag}/M \simeq 32$ respectively, i.e indicating the
presence of a constant lag in this particular phase too. In order to
get a better understanding of the orbits responsible for these peaks
and their relation to the corresponding orbits of the Kerr geometry
we did obtain and plot in Figure~\ref{fig:ray-a0} the orbits of the
photons corresponding to these peaks. Thus the largest peak
corresponds to photons that arrive at the observer directly from the
source (in the {\sl retrograde} direction with respect to the source
direction), while the smaller one from photons that go around the
hole by an angle $\Delta \phi \simeq 3\pi/2$ in the {\sl prograde}
direction. The next much smaller peak at $t/M = 632 + 16$ is due to
photons that arrive at the observer after going around the hole by
an angle $\Delta \phi \simeq 5\pi/2$ in the {\sl retrograde}
direction and the one at $t/M = 632 + 32$ after an angle $\Delta
\phi \simeq 7 \pi/2$ in the {\sl prograde} direction. As the
azimuthal angle $\phi_{s,i}$ approaches 180$^{\circ}$, both the two
largest and two smallest peaks move closer (as expected by symmetry)
to obtain approximately the same normalization and merge for
$\phi_{s,i} \simeq 180^{\circ}$, indicating that their time-lag is
in fact not constant but depends strongly on the value of the phase
angle $\phi_{s,i}$, unlike the situation for the Kerr geometry of
Figure~\ref{fig:response-a099}, where the lags are independent of
the source phase. This is a crucial point in our model.

\section{Model X-ray Light Curves and Timing Analysis}

We have used the response function constructed as above to produce
synthetic light curves. In order to avoid the introduction of QPO
features by the rotation of the emitting sources, we have made the
flare emission instantaneous and positioned the sources at {\sl
random} values of $\phi_{s,i}$, uniformly between 0 and $2 \pi$; we
also introduced the flares randomly in time using the following
prescription of the time interval $\Delta T_i$ between the $i$-{\rm
th} and $(i+1)$-{\rm th} bursts
\begin{eqnarray}
\Delta T_i \equiv \bar{T} \times \vert \ln \left\{ \rm{rnd}\left(0,1
\right)\right\} \vert \ , \label{eq:time}
\end{eqnarray}
where $\bar{T}$ is a mean timescale between bursts and
$\rm{rnd}(0,1)$ denotes a random number between $0$ and $1$. We have
chosen $\bar{T} \equiv f T_{\rm orb}$ where $T_{\rm orb}(r_s) \equiv
2 \pi (r_s^{3/2}+a M^{1/2})/M^{1/2}$ is the Keplerian orbital period
at $r=r_s$ and $f$ ($f \sim 1$) characterizes the frequency of burst
occurences relative to the orbital period $T_{\rm orb}$ (the precise
value of $f$ is in fact not relevant for the shape of the power
spectrum we are interested in).

Each burst produces a characteristic signal $I_i(t;r_s,\phi_{s,i})$
unique to that specific azimuthal position $\phi_{s,i}$ with
$I_i(t;r_s,\phi_{s,i})$ denoting the intensity of the signal of the
$i$-{\rm th} burst as shown in Figures~\ref{fig:response-a099} and
\ref{fig:response-a0}. The entire bolometric light curve then is the
incoherent superposition of similar signals corresponding to the
randomly produced flares, i.e.
\begin{eqnarray}
I(t;r_s) \equiv \sum_{i=1}^{N_b} I_i(t;r_s,\phi_{s,i}) \ ,
\label{eq:lightcurve}
\end{eqnarray}
where $N_b$ (set to be 6,000) is the total number of X-ray bursts
used in a given model light curve. Note that each flash contains
$N_{\rm ph}=200,000$ photons.




\begin{figure}[t]
\epsscale{0.99} \plottwo{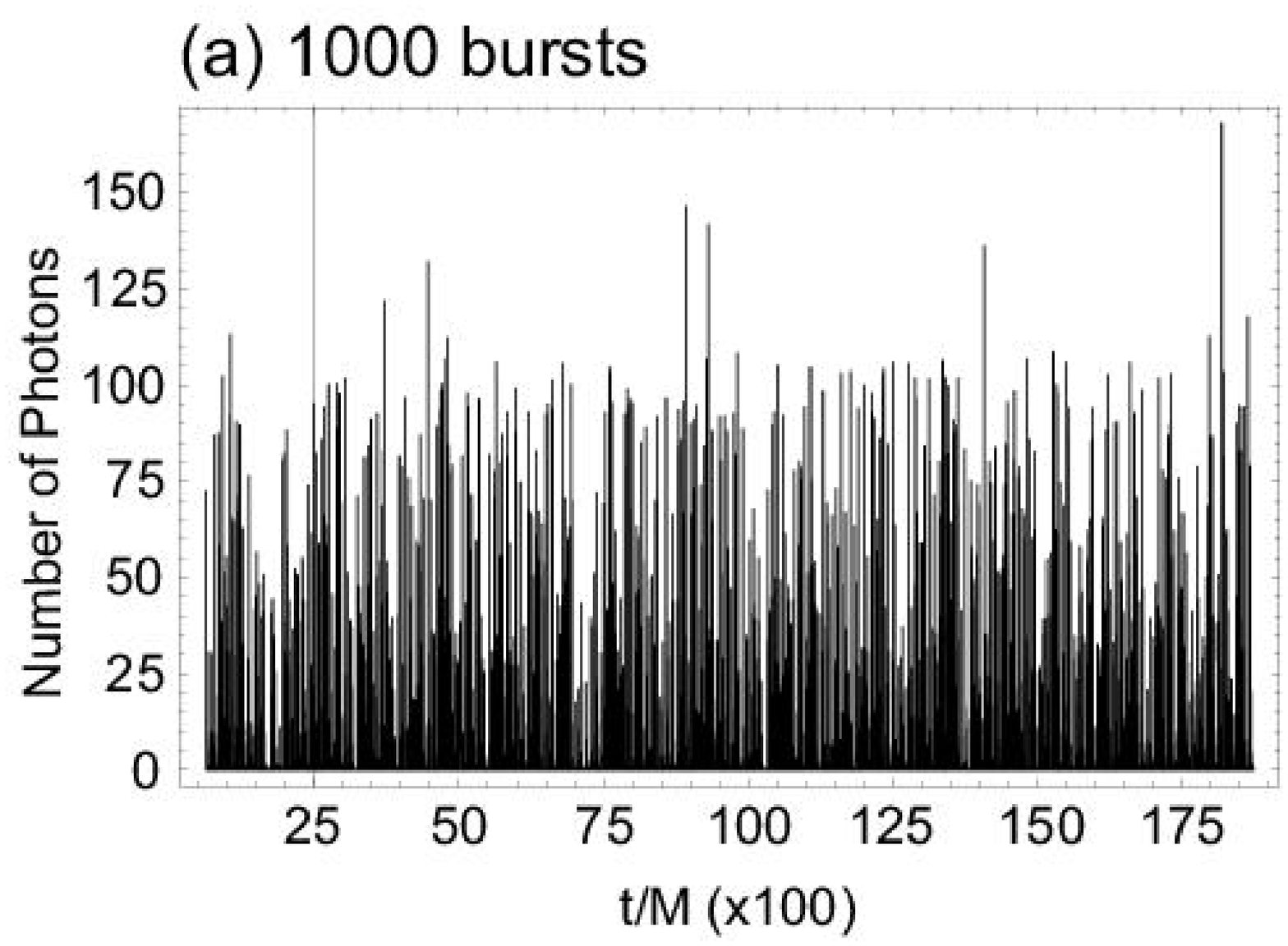}{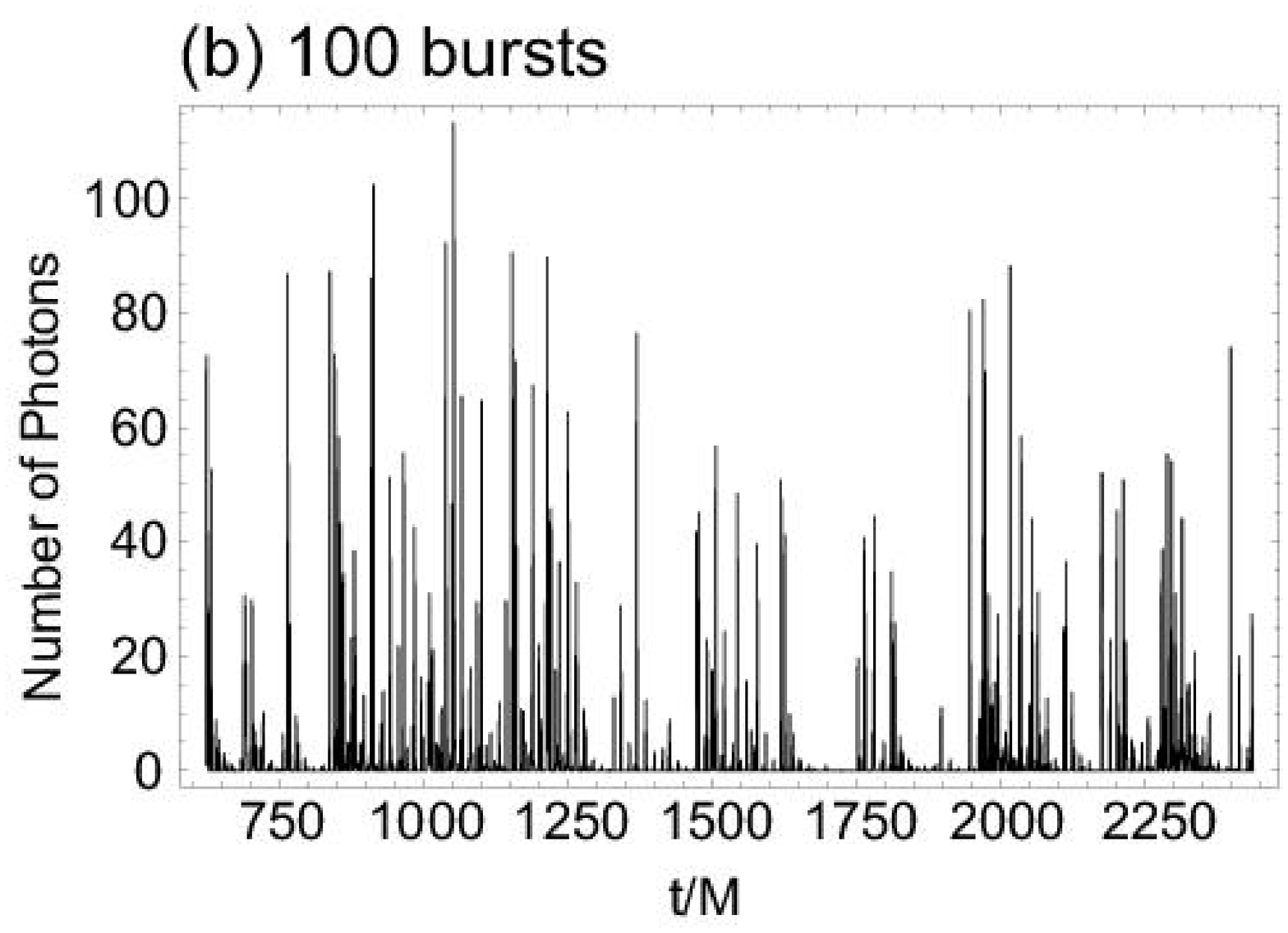} \caption{Two sample light
curves generated as discussed in the text. (a) The light curve of a
series of $N_b=1000$ bursts from sources at radius $r = r_{\rm ms} =
1.45 M$ for a black hole with $a/M=0.99$ and $f=1$. (b) The light
curve of only the first 100 bursts of the same series. It is
apparent that these are totally random. \label{fig:lightcurve}}
\end{figure}

Using the above prescription we have produced sets of model light
curves varying the parameters $N_b$ and $f$ for the same values of
the black hole parameters used to produce Figures
\ref{fig:response-a099} and \ref{fig:response-a0}.
The simulated light curve seen by the observer is shown in
Figure~\ref{fig:lightcurve} where $N_b=1000$ bursts are considered
in (a) and the first 100 bursts are extracted from (a) in (b). They
exhibit no apparent periodicity, given the random positions
$\phi_{s,i}$ and times of the induced flares $\Delta T_i$. We find
that the characteristic appearance of the light curve essentially
remains the same for the \sw cases too.

\begin{figure}[t]
\epsscale{0.45} \plotone{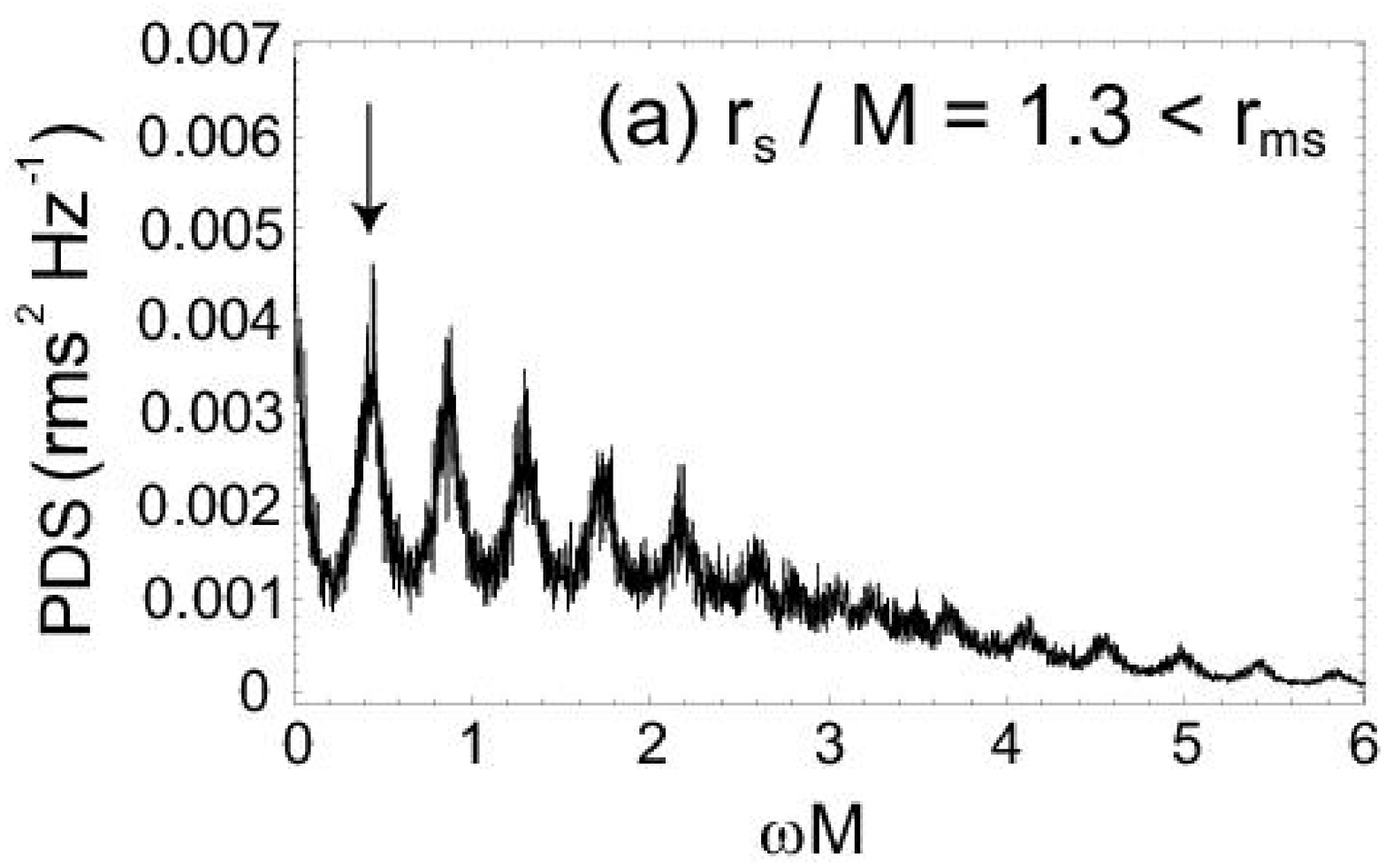}\plotone{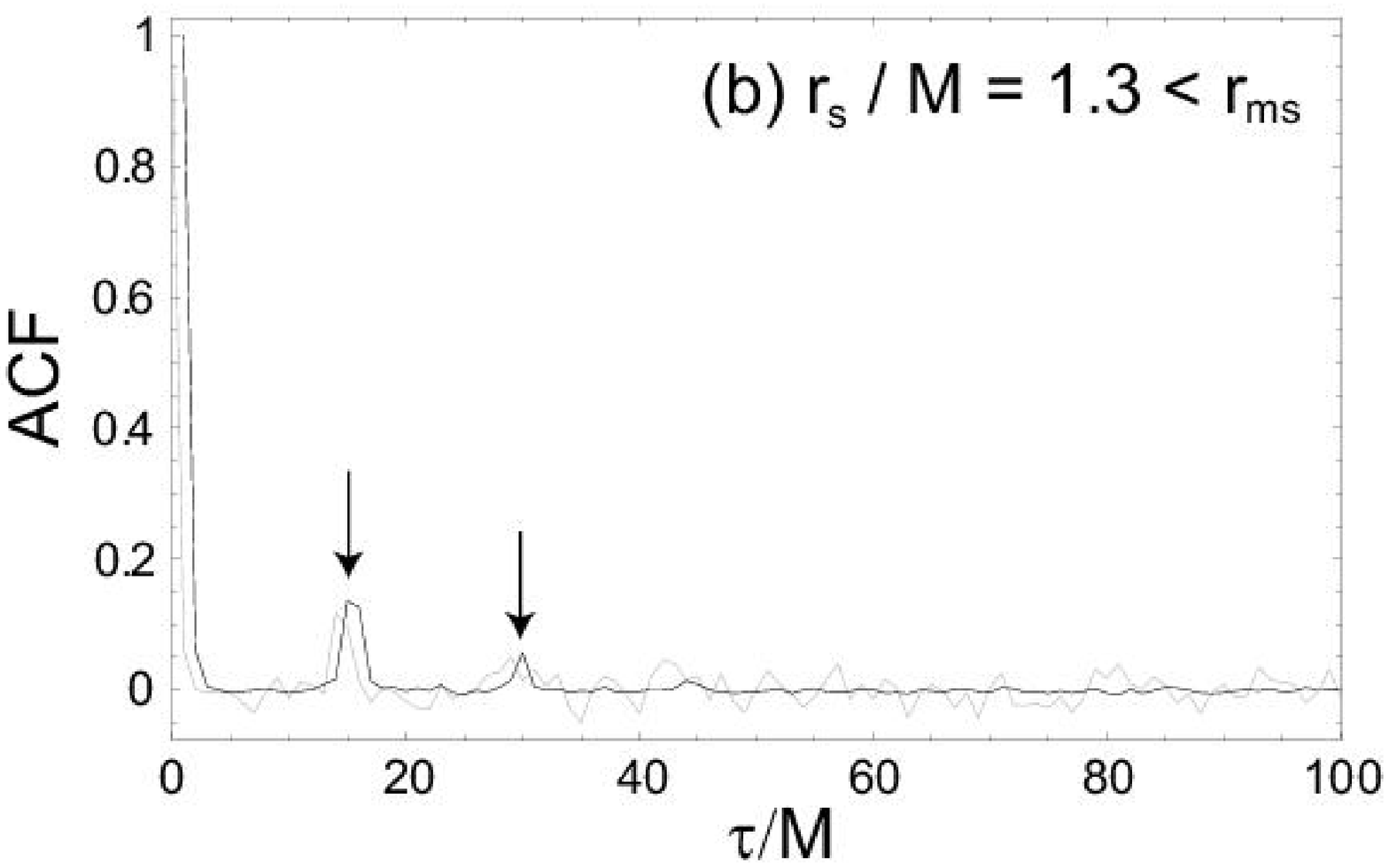}
\plotone{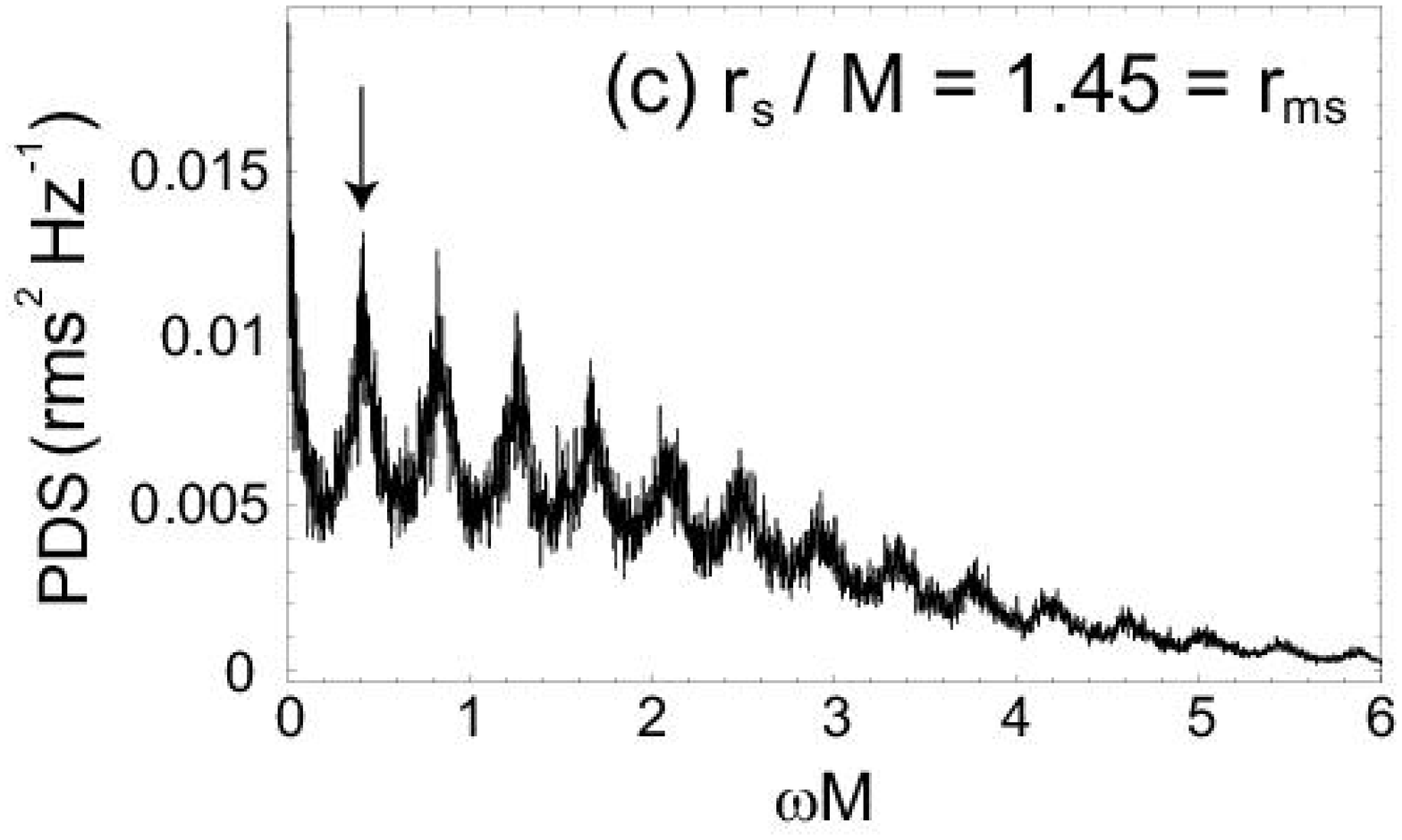}\plotone{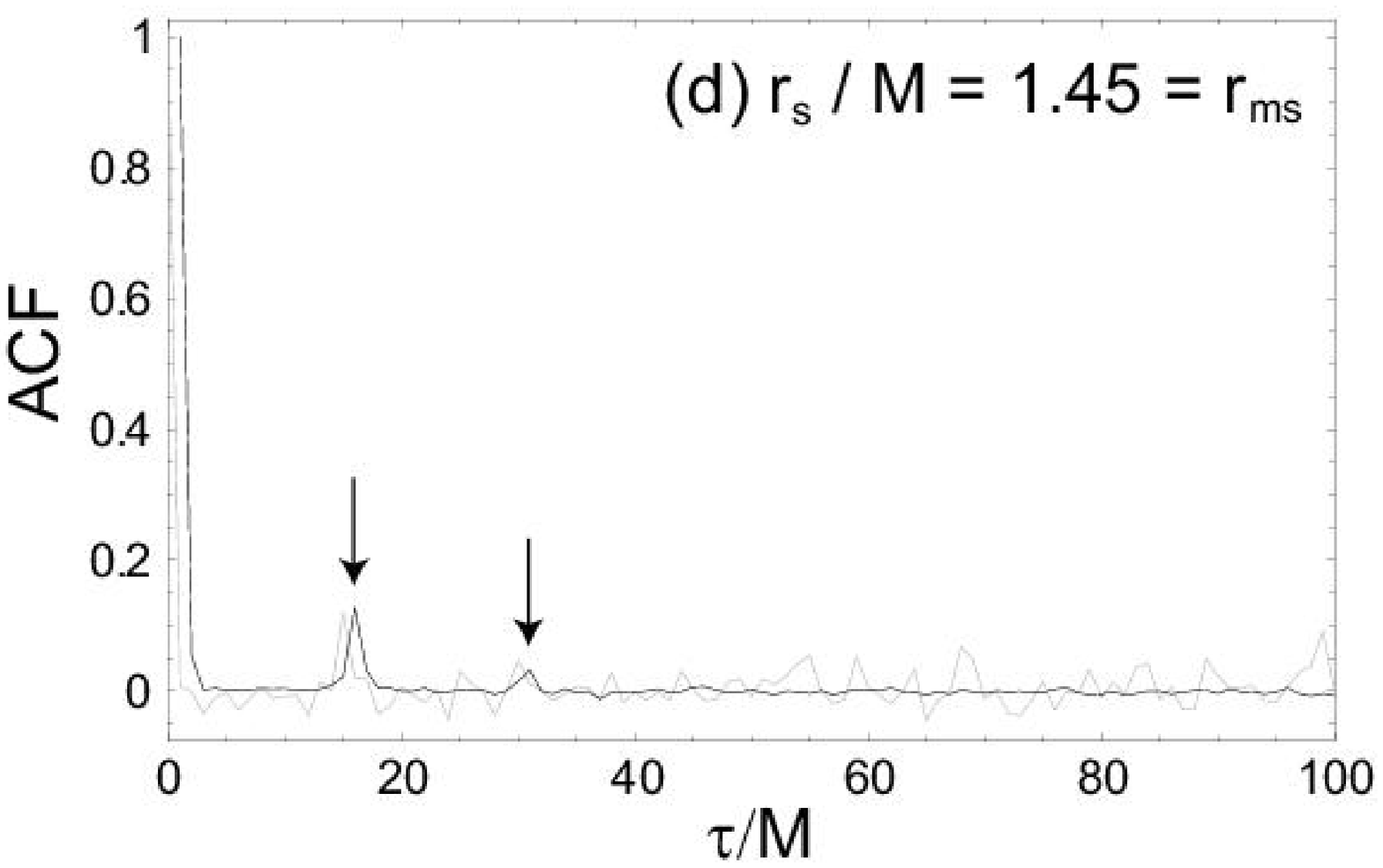}
\plotone{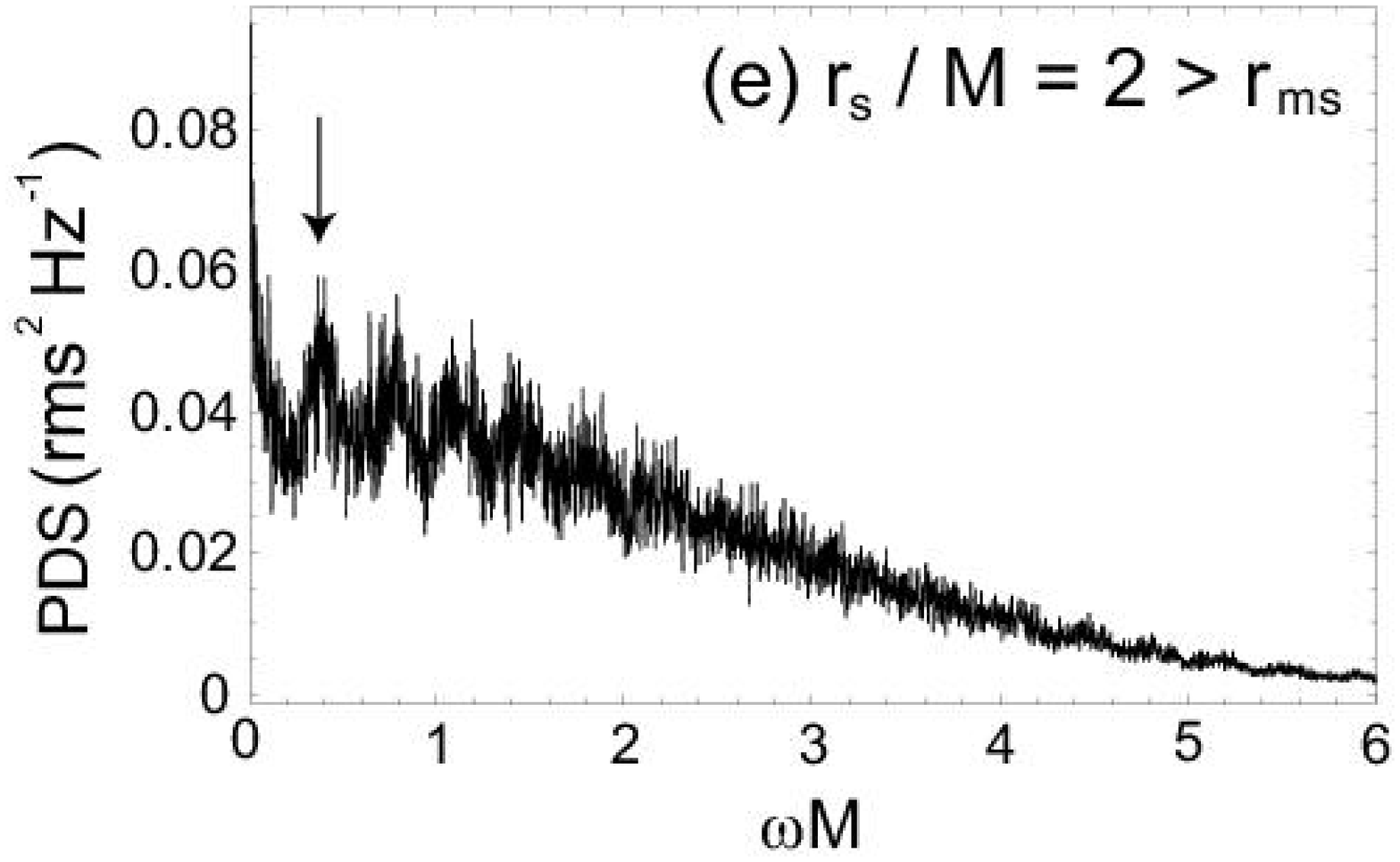}\plotone{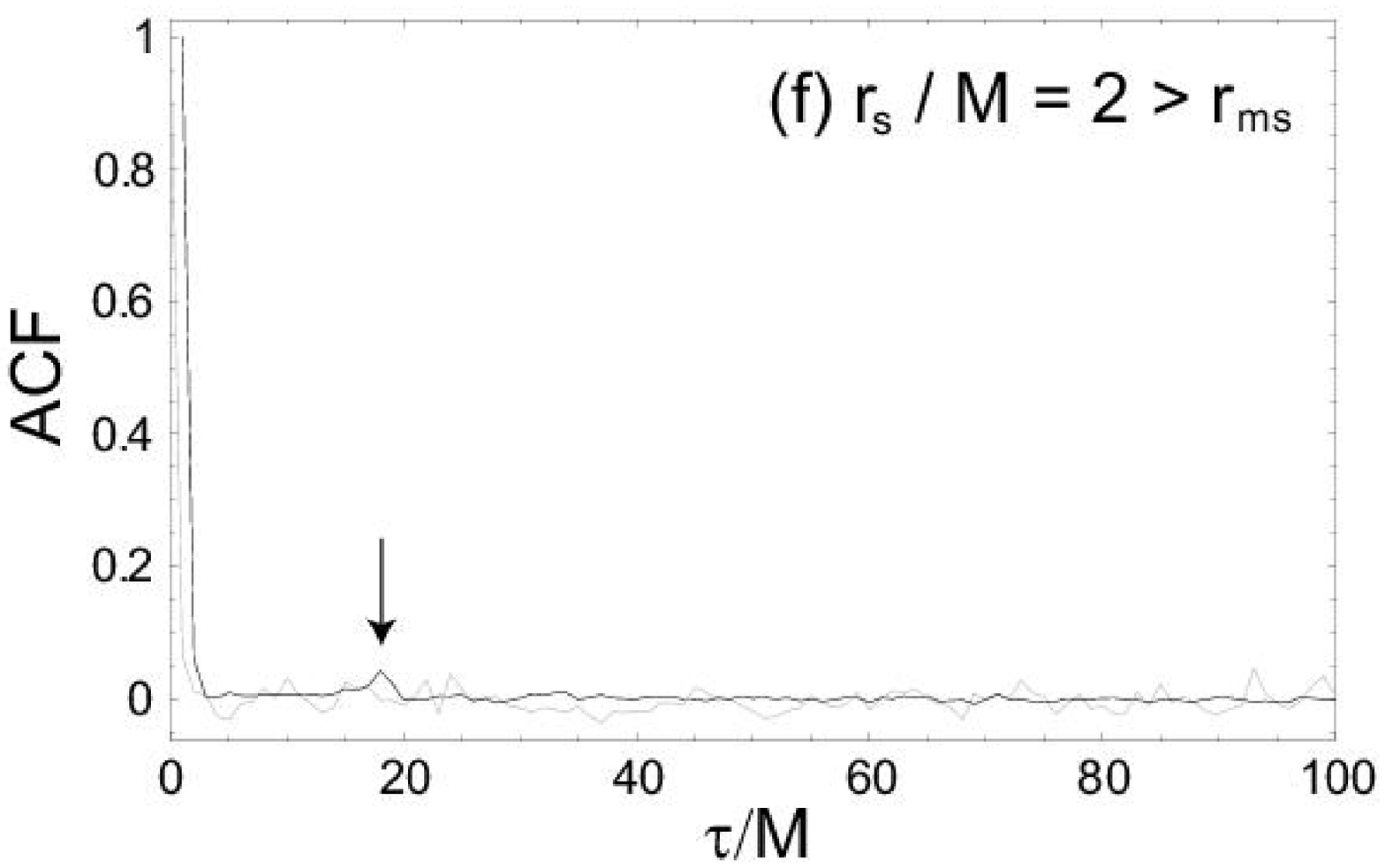} \caption{Timing analysis of
random X-ray bursts with $a/M=0.99$ for $r_s/M=1.3$ (top rows),
$1.45$ (middle rows) and $2$ (bottom rows). We set $N_b=6000$. Left
columns show the average power density spectra (PDS) while right
columns show the corresponding autocorrelation functions (ACF). In
the ACF we show two cases: $N_b=6000$ (dark curves) and $100$ (gray
curves) for comparison. \label{fig:PDS1}}
\end{figure}

We have then used standard analysis tools for the Fourier transform
$F(\omega)$ to produce the PDS $\vert F(\omega) \vert^2$ of these
light curves along with the Autocorrelation Function (ACF)
$R(\tau)$. These are shown in Figure~\ref{fig:PDS1} for various
source radii, i.e. $r_s > r_{\rm ms}, ~r_s=r_{\rm ms}$ and { a
plunging orbit with} $r_s < r_{\rm ms}$, for a black hole with
$a/M=0.99$. A characteristic peak is clearly seen in the PDS at
angular frequency of $\omega_{\rm QPO} M \sim 0.4$ and its higher
multiple modes, which for $M=10\Msun$ corresponds to a peak
frequency $\nu_{\rm QPO} \sim 1.3-1.4$ kHz. The ACF exhibits also a
prominent peak at the equivalent time lag of $\tau/M \sim 14 - 15$
and perhaps encapsulates better the underlying physics behind the
QPO, i.e. the presence of a well-defined time lag (i.e. an ``echo")
in the response of this system. This is best seen, e.g., in
Figures~\ref{fig:PDS1}a-d. The QPO pattern (peak frequency and
width) remains essentially the same regardless of the source
position, as long as it is within the ergosphere. On the other hand,
as $r_s$ approaches the static limit ($r=r_{\rm static}=2M$ near the
equator), the QPO signal begins to disappear (see
Figs.~\ref{fig:PDS1}e and f), since the time-lags between the peaks
of the response function is no longer constant, as discussed
earlier.

It should be noted that the choice of the times of flare injection
[equation~(\ref{eq:time})] guarantees that in the absence of the
lags discussed above the corresponding PDS would be just white noise.
However, because our timing resolution in this numerical experiment
is $M$, we have approximated each ``bunch" of photons received at a
given angle and a given resolution interval as a Gaussian of area
normalized to the number of photons received in each interval and
FWHM equal to $M$. It is precisely this finite width of our ``shots"
that leads to the high frequency cut-off of the PDS given in
Figures~\ref{fig:PDS1}a-d. Of course, at the low frequencies the PDS
is that of white noise.

\begin{figure}[t]
\epsscale{0.45} \plotone{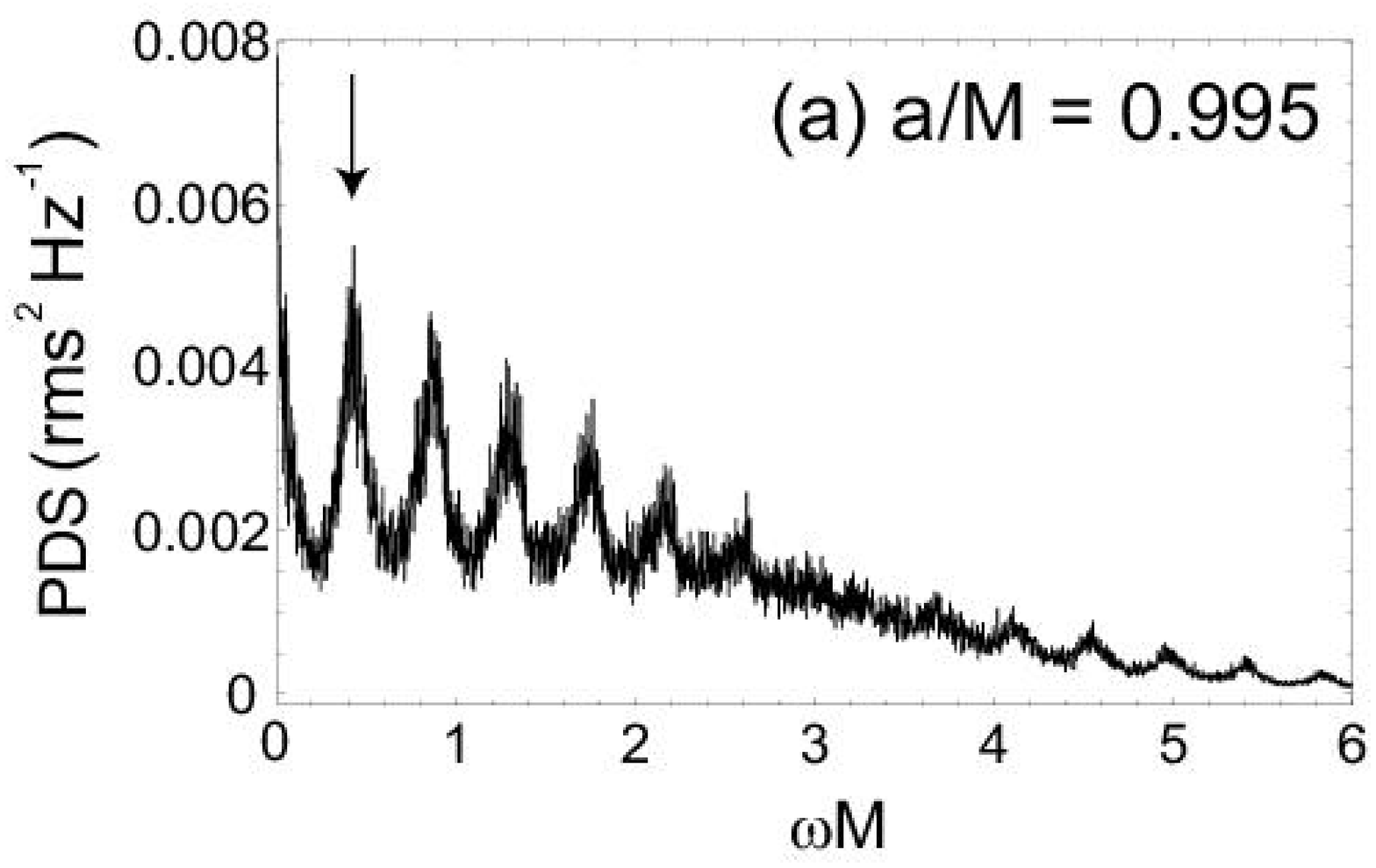}\plotone{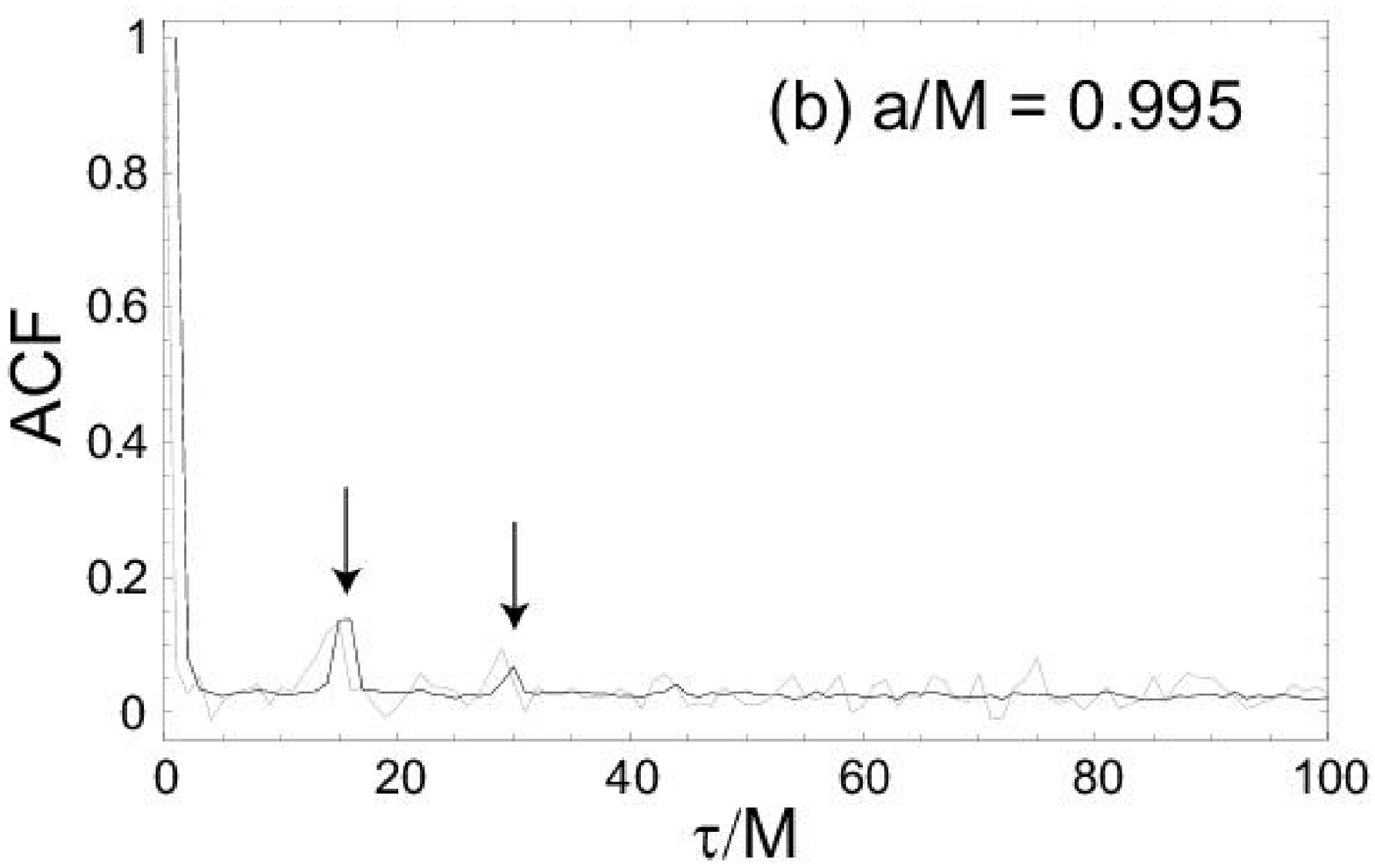}
\plotone{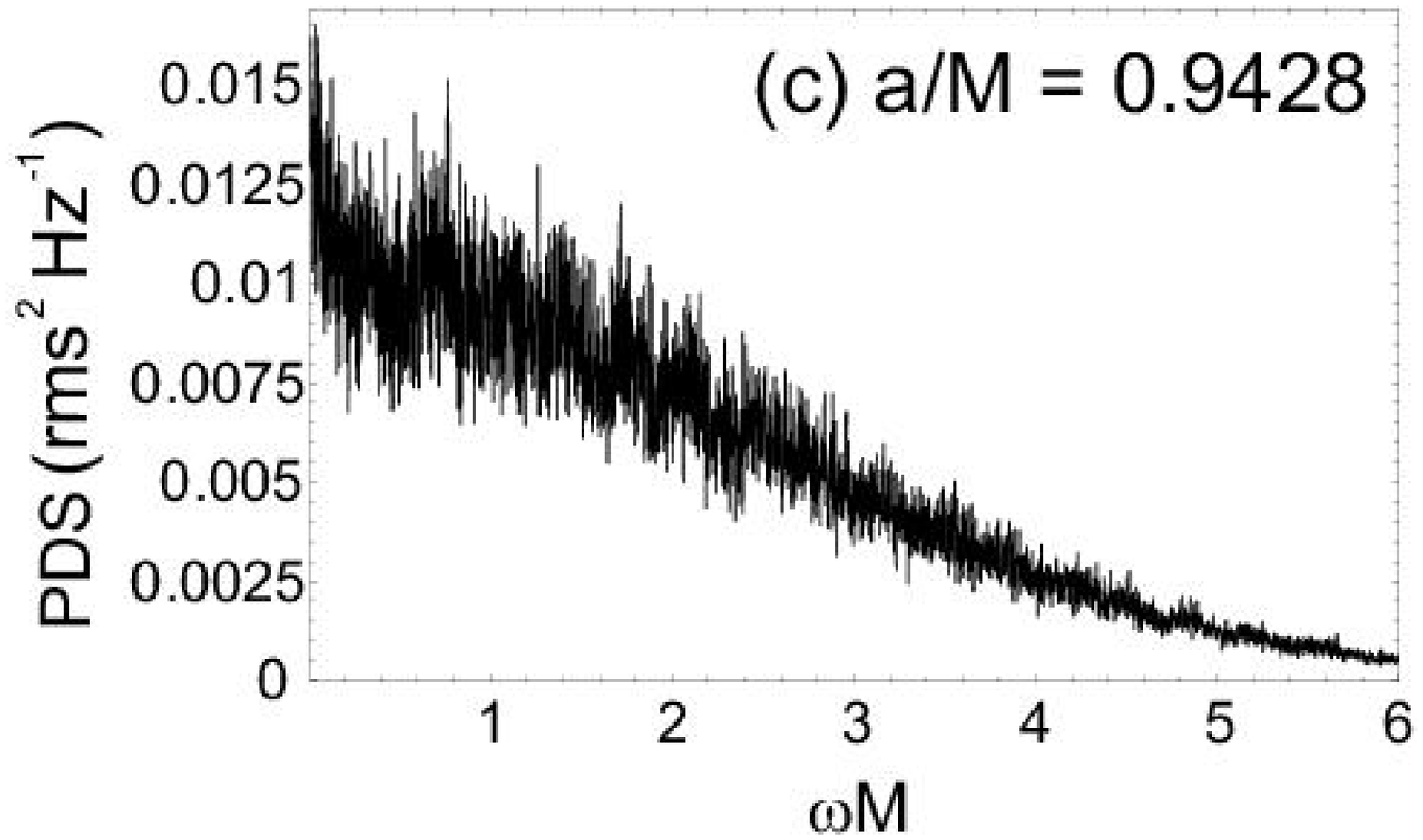}\plotone{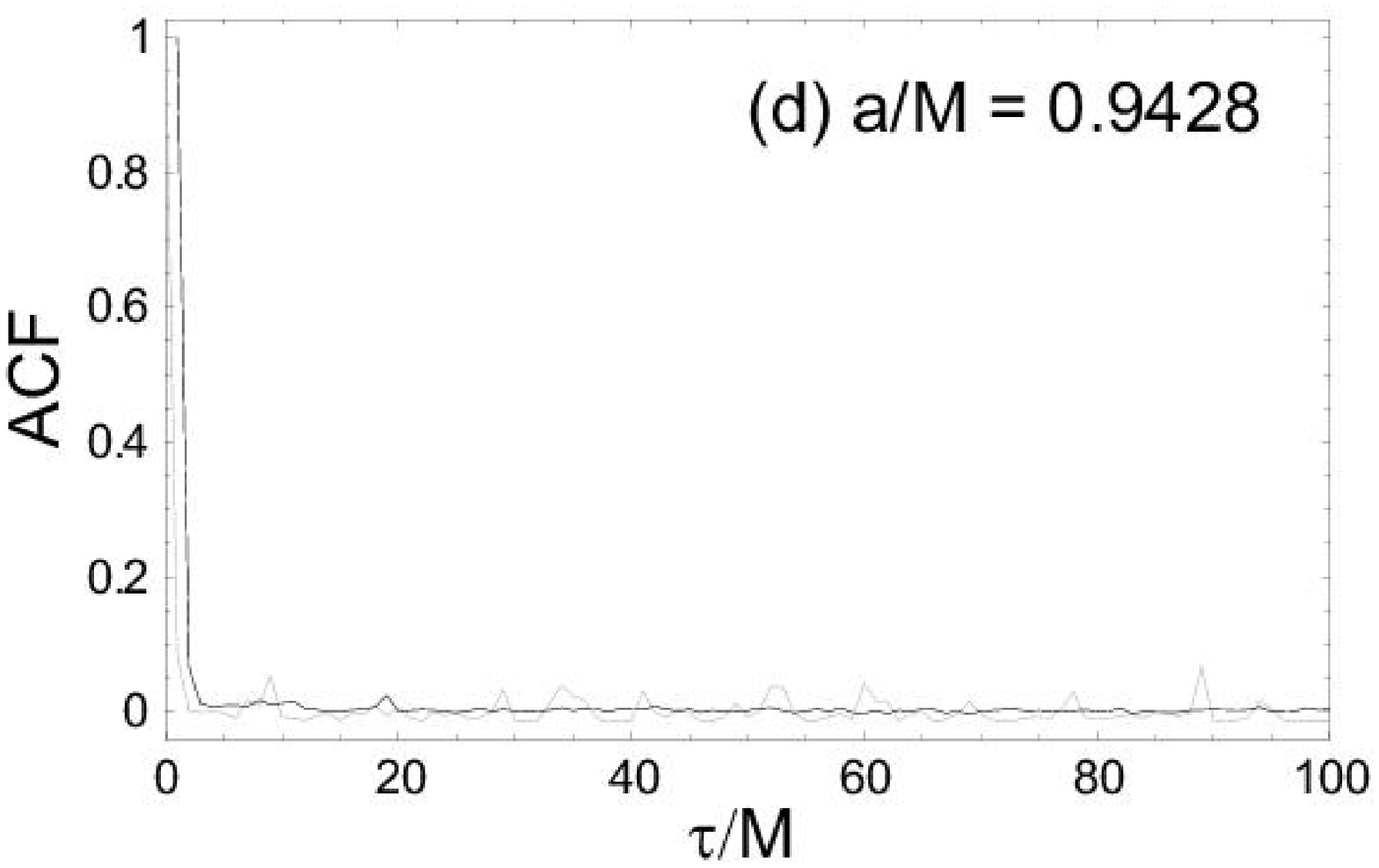}
 \caption{Same as Figure~\ref{fig:PDS1} but for $a/M=0.995$ (top columns)
 and $0.9428$ (bottom columns) with $r_s=r_{\rm ms}(a)$. \label{fig:PDS2}}
\end{figure}

In order to study quantitatively the QPO feature in this model we
have also examined its dependence on the black hole rotation $a$ for
$r_s=r_{\rm ms}$ in Figure~\ref{fig:PDS2}. To do so we select two
values for the black hole spin, namely $a/M=0.995$ and $a/M=0.9428$.
The first was chosen to verify that the effect becomes more
prominent with the increase of the black hole spin; the second value
was chosen by the requirement that the equatorial boundary of the
ergosphere be equal to that of the ISCO, i.e. that $r_{\rm ms} =
r_{\rm static}=2M$. The first case yields clearly prominent QPO and
peaks at the appropriate lags in the ACF. However, these features
become almost statistically insignificant as the spin parameter
approaches the critical value of $a/M=0.9428$ (for which $r_s=r_{\rm
ms}=r_{\rm static}$) for the reasons discussed in the previous
paragraphs of this section.

We have also performed a similar calculation for a \sw black hole case
for $r_s=r_{\rm ms}=6M$ to examine whether the QPO features we have
found in Kerr geometry may also be produced in the absence of black
hole rotation.
%
%
In this case the PDS (not shown here) exhibits no QPO features
similar to those of the Kerr case (as expected). The reason is the
absence of a constant time-lag between the two major peaks of the
response function (see Fig.~\ref{fig:response-a0}); the
normalization of the two minor peaks is just too small to make a
difference in the PDS, despite the presence of roughly constant lags
between them at some phases. The corresponding ACF also showed no
particular coherent timescale. We have repeated similar computations
for various source radii ($r_s \gtrsim r_{\rm ms}$ and $r_s < r_{\rm
ms}$) for \sw case and found that in no case the QPO features seen
above were produced. As discussed above this is due to the absence
of a constant time lag in the system response. We will come back to
discuss this issue in \S 4. Of course, should there be a reason that
a particular phase is favored as the site of the X-ray flares
considered by our model, then one would expect QPOs to be present
for sources in a \sw geometry too.

Finally, we have checked that the feature obtained is not simply an
artifact due to photon statistics by changing the number of bursts
and the random number sequence used and indeed nearly identical
results were consistently found. Note that for a given black hole
rotation $a$, there are no free parameters other than the radial
position of the flares $r_s$, and the mean timescale of each burst
$\bar{T}$, neither of which changes the value of the lag in the
response function.

%
%
%
%

\section{Discussion}

We have presented above a process that can lead to QPO features in
the PDS of accreting black hole systems. This process is to a
certain extent different from the typical QPO models in the
literature in that it relies on the ``echo" of a (however random)
signal for the emergence of the QPO in the observed PDS. By
contrast, most QPO models invoke the presence of an underlying (but
perhaps obscured due to noise) oscillation in the emitted photon
flux. Whether this qualifies it as an alternative QPO model or not
is, in our view, a matter of convention. However, it provides a
mind-set concerning the QPO phenomenon that deviates from that of
the more conventional considerations which may lead to novel future
insights on the subject. The difference of these two distinct QPO
notions is perhaps exemplified best mathematically by the
corresponding ACF, which in one case has an oscillatory behavior
while in the other the double peak form presented in
Figures~\ref{fig:PDS1} and \ref{fig:PDS2}.

As discussed in the previous section, the ``echo" of the proposed
model is the result of the black hole angular momentum $a$ and the
ensuing dragging of inertial frames on the trajectories of the
photons emitted within its ergosphere; as such,  this ``echo" is
absent in black holes of sufficiently small values of $a$. This is
not the first time that frame-dragging has been invoked to account
for a certain aspect of the QPO phenomenon. For example, the same
effect and the accompanying  Lense--Thirring disk precession has
been invoked to account for the dependence of the intrafrequency
correlations of several low frequency QPOs \citep{Stella98, Klis00,
Schnittman06}; the effect we describe herein is different in that it
affects the orbits of individual photons, rather the orientation of
the entire disk.

One should further note that, because the ``echo" lags that produde
these QPOs depend only on the background geometry of the accreting
black hole (they are roughly equal to the length of the circular
photon orbit for this geometry), they (and also the resulting QPO
frequencies) are independent of the source flux (the accretion
rate). On the other hand, given that the QPOs of accreting neutron
stars and most of the QPOs of accreting black holes do depend, in
general, on the source flux \citep{Klis00}, they are likely due
processes different from that described above. Nonetheless, as
discussed in \S 1, QPOs that do not depend on the source flux have
been discovered in a number of galactic black hole candidates
\citep{Strohmayer01a, Strohmayer01b}. However, the frequencies of
these latter QPOs are much too small ($\nu_{\rm QPO} \sim 100 - 400$
Hz) to be attributed to the process described herein [$\nu_{\rm QPO}
\simeq 1400 \, (10M_{\odot}/M)]$, unless our understanding of strong
gravity physics is in significant error.

Up to this point, the discussion of our model considered flares
whose duration is much shorter than the orbital period of the
accretion disk near its ISCO. Clearly, the ``echo" process discussed
herein is present for longer flare durations, even if the latter
exceeds the local orbital period. However, in this case the model
becomes very similar to those of orbiting hot spots discussed
earlier \cite[][]{Karas99, Schnittman05}. We have produced a number
of model light curves assuming the flare duration to be several
times the ISCO orbital period (i.e. just like in the orbiting hot
spot model) and computed the corresponding PDS and ACF. In this
case, the geometrical echo, while still present, makes only a small
contribution to the PDS, whose QPOs are now dominated by the
orbiting hot spot periodic motion.

While the process we discussed above is quite robust (it depends
only on the geometry of the accreting object), its potential
observability depends on a number of factors. To begin with, our
calculations were performed exclusively on the equatorial plane of a
Kerr black hole; therefore, our present results are valid only for
disks that are geometrically thin and for observers at relatively
low latitudes. For observers at moderately higher latitudes (or
thick disks), our orbit calculations have to be supplemented with
those for the $\theta-$coordinate (to follow the poloidal motion)
that have been omitted so far in this work. However, the essence of
the effect we discussed above, i.e. forward thrust of all photon
trajectories produced within the ergosphere, whether equatorial or
not, is expected to be present in that case too and thus
qualitatively we expect the same phenomenon to be conditionally
observable for non-equatorial source/observer configurations. Our
preliminary calculations indicate the presence of such QPO for
observers at latitudes at least as high as $\sim 30^{\circ}$ from
the disk mid-plane. Quantitative analysis of this aspect of the
problem, which bears the application of these ideas to thick disks,
will be presented in a future publication.


Another limitation of the model discussed in the earlier sections is
that of the source radius $r_s$. The response frames of
Figure~\ref{fig:response-a099} were computed assuming the source to
be on the ISCO at $r_s/M \simeq 1.45$.
Changing the source radius leads to a quantitatively similar
behavior, i.e. the ACF still exhibits a peak at the same value of
the lag $\tau/M \simeq 14$, as long as the source of photons is
located within the static limit (ergosphere), i.e. for $r_s/M \lsim
2$ near the equatorial plane. Interestingly, for the value $a/M =
0.99$ used in these calculations the ACF peak at $\tau/M \simeq 14$
disappears gradually as the source radius increases past $r_s/M
\gsim 2$, i.e. as the source moves outside the ergosphere, a fact
which suggests that the effect we have presented is in fact due to the
strong dragging of inertial frames. Equivalently, we anticipate the
gradual disappearance of this peak in the ACF as the value of the
black hole spin $a$ decreases down to a critical value of $a/M \sim
0.9428$ (where the radius of ISCO becomes equal to that of static
limit) and the ergosphere shrinks to leave much of the disk outside.
While the frame-dragging is still operative even at radii outside
the static limit, it is simply not effective enough to cause
significant azimuthal beaming.

We have in fact checked that the effect we consider is not due to
the source proximity to the horizon (a situation possible without
free fall onto the hole only if the latter is rapidly rotating) by
computing the orbits of photons from radii as small as $r_s/M \simeq
2.1$ by matter in-falling (in the plunging region of $r_s < r_{\rm
ms}$) onto a \sw black hole. In computing the photon orbits in this
last case we have taken into account all the components of the
velocity of the in-falling plasma, calculated by assuming that it
began its in-fall from the ISCO. Plunging gas thus preserves its
ISCO energy and angular momentum. We found that for flares that take
place close to the horizon ($r_s/M \sim 2.1$) most photons end up
into the black hole (as expected); however, the behavior of the
response of the escaping photons in no case gave us the constant
time-lag obtained in the rapidly rotating black hole case, to
produce QPO features in the PDS. In the case of a \sw black hole, an
(unstable) photon circular orbit lies at $r_s=3M < r_{\rm ms}$
inside the plunging region, from which radius one would expect
higher-order photons (multiple orbit photons). However, since the
plunging X-ray sources also possess significant radial velocity
component at this radius close to the horizon, many photons emitted
at $r_s/M=3$ are Doppler-beamed in the direction of the source
motion. The radial beaming effect therefore keeps most of the
photons from being emitted in an azimuthal direction, effectively
suppressing their multiple orbits.

\begin{figure}[t]
\epsscale{0.99} \plottwo{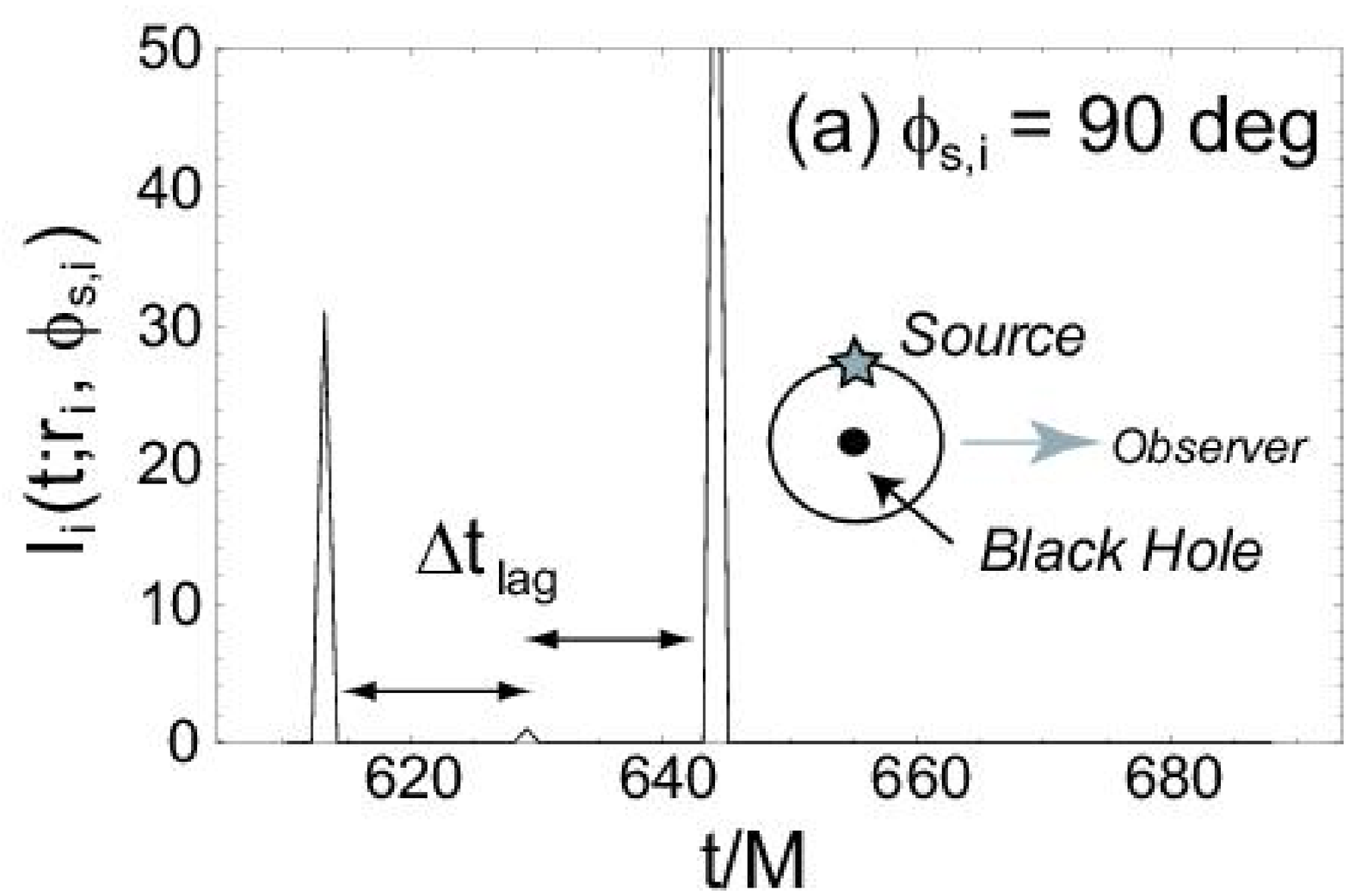}{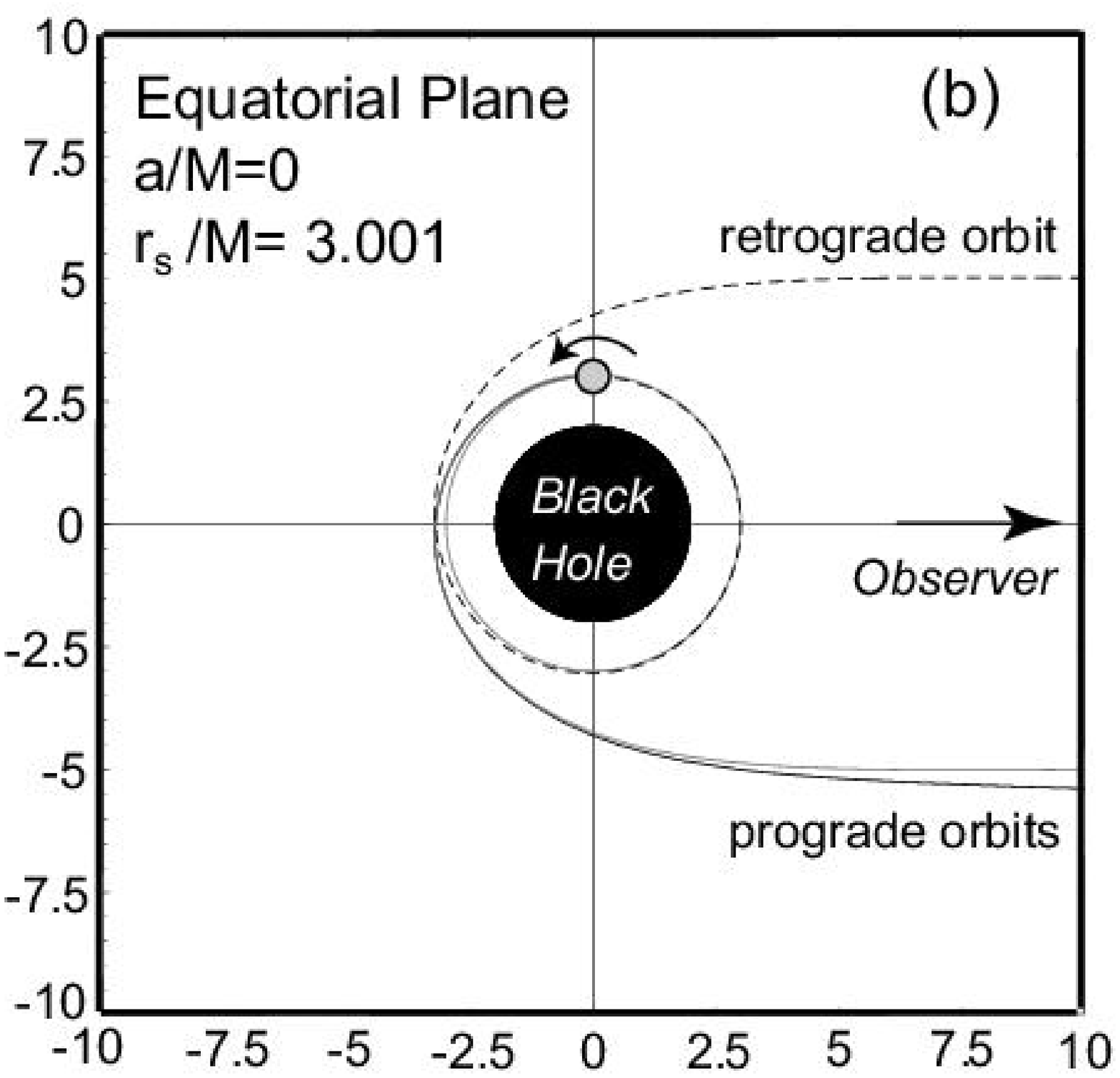} \caption{(a) The response
function for a source at an unstable (Keplerian) circular orbit at
$r_s/M=3.001$ at a phase $\phi_s=90^{\circ}$. The two large peaks
correspond to the prograde orbits that cover angles $\Delta \phi =
\pi+\pi/2$ and $3 \pi+\pi/2$ respectively. The small peak at
midpoint corresponds to the retrograde orbit that covers an angle
$\Delta \phi = 2 \pi+\pi/2$. (b) The photon orbits corresponding to
the features of the figure on the left. \label{fig:r3m-ray}}
\end{figure}

\begin{figure}[t]
\epsscale{0.99} \plottwo{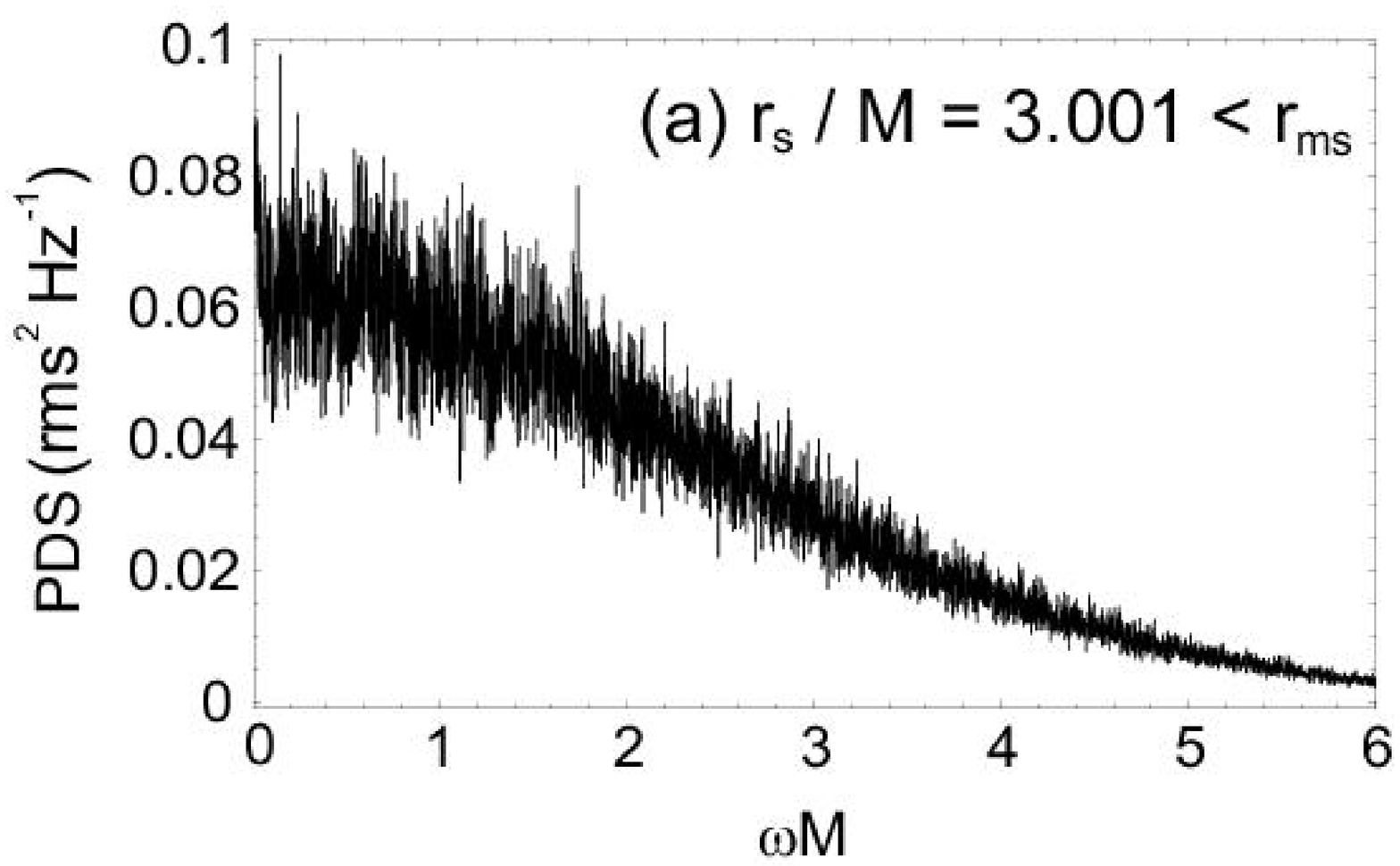}{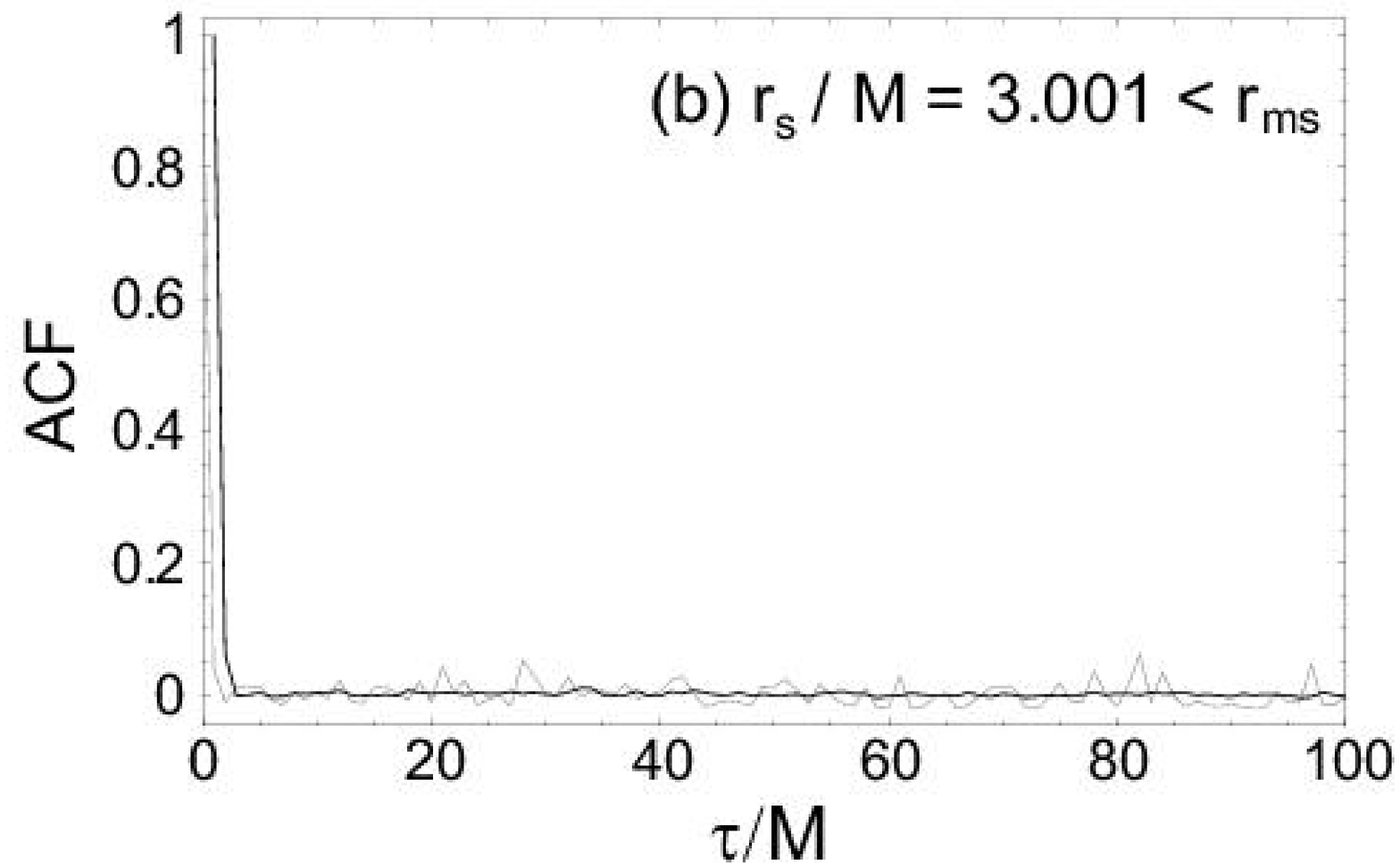} \caption{Predicted PDS of
the light curve of a source at $r_s/M = 3.001$ obtained using the
response function of Figure~\ref{fig:r3m-ray} and the prescription
given in the text. (b) The ACF of the same light curve for
$N_b=6000$ (dark curves) and $N_b=100$ (light curves).
\label{fig:r3m-respds}}
\end{figure}

Finally, to explore even the most likely case to produce QPOs in a
\sw geometry, we computed the response function of the system for a
source at an unstable (Keplerian) circular orbit at $r/M = 3.001$.
At this radius, the high rotational speed of the source ($v \simeq
0.9995 c$) produces effects similar to the dragging of inertial
frames, because most photons are beamed in the forward direction by
the source rotation. The corresponding response function and photon
orbits for a source at $\phi_{s,i} \simeq 90^{\circ}$ are given in
Figures~\ref{fig:r3m-ray}a,b, while the corresponding PDS and ACF in
Figures~\ref{fig:r3m-respds}a,b. As it can be seen there, for this
specific source position, most photons reach the observer after
going around the black hole by $\Delta \phi \simeq 3 \pi+\pi/2$
while the fraction that reaches after $\Delta \phi \simeq \pi
+\pi/2$ is comparable but smaller; there are also a small number of
photons that reach the observer in the counter clockwise direction
after $\Delta \phi \simeq 2 \pi+\pi/2$; photons that reach the
observer after a larger number of rotations are not discernible at
this resolution. However, as the source phase $\phi_{s}$ changes,
while the lag between the major peaks remains roughly constant, the
relative normalization shifts quickly and for most of phases the
response is dominated by a single peak. This fact, along with the
random choice of the phase of a given burst of our light curve
prescription, ``washes-out" potential QPO features, leading to the
PDS shown in Figure~\ref{fig:r3m-respds}a. On the contrary, in the
Kerr case, the frame-dragging aided azimuthal beaming helps a
significant fraction of the photons orbit around the black hole for
all values of the phase angle $\phi_{s}$. This is an essential
difference between \sw and Kerr black hole cases, which we conclude
is manifested as the QPO feature we see here.

Up to this point we have only considered the time correlations
between the photons emitted by the accretion disk without any
reference to their energy. However, if these photons are of a
specific energy (i.e. the Fe transitions observed in accretion
disks), there exists a relation (redshift) between the emitted and
the received photon energies given by
\begin{eqnarray}
g \equiv \frac{\nu_{\rm rec}}{\nu_{\rm emt}} = \frac{\left(r_s^2
\Delta_s / A_s \right)^{1/2} \left(1-v_s^2 \right)^{1/2}} {\left(1-b
\Omega_K \right)} \ , \label{eq:redshift}
\end{eqnarray}
where $\Omega_K$ is the Keplerian angular frequency, and $v_s$ is
the azimuthal component of the X-ray source velocity measured in
LNRF. It would therefore be of interest to consider, in addition to
the time correlations, also their additional dependence on the
photon energy (or energy shift factor) $g$; the impetus for such a
study comes from the observed energy dependence of the geometry
attributed QPOs in GRO~J1655-40 \citep[e.g.][]{Strohmayer01a}. We
hope to look into this as well as the above unresolved issues in a
future publication.

We would like to conclude with some general remarks concerning the
``light echo" responsible for the QPOs presented in this paper: This
is a very generic process and would be present at any source that
exhibits a delay in its response function, not necessarily the one
we described in this note. The same point has been made earlier
\citep{KazHua99} in a different context and we believe that it may
have broader applicability. It should be noted, in this same
context, that the response of an axisymmetric disk does not exhibit
any such features, however, the response of a {\sl warped disk}
\citep[e.g.][]{Hickox05} should; therefore such disks may exhibit
QPO features of the kind described herein at the periods comparable
to the light crossing time across the disk, features that maybe
worth searching for in the data.

Finally, we would like to point out that the lack of apparent phase
coherence in this model makes the PDS quite noisy and therefore we
expect that the detection of such features may require their search
in the light curves of higher mass objects that would push the
corresponding frequencies to lower values that are easier to detect
(perhaps intermediate mass black holes associated with ULXs or
nearby AGN), using low background, high throughput missions like
{\it Constellation-X}.

\acknowledgments

We would like to thank the anonymous referee for a number of useful
and insightful suggestions. This research was supported in part by
an appointment to the NASA Postdoctoral Program at the Goddard Space
Flight Center, administered by Oak Ridge Associated Universities
through a contract with NASA and an INTEGRAL GO grant.

\section{Appendix}

In this Appendix we show the definition of the local angle $\delta$
of photon emission in LNRF and its relation to photon's impact
parameter $b$ (which is defined as specific angular momentum). We
then explicitly derive radial component of the geodesic equation in
terms of $\delta$ which we have integrated in producing
Figures~\ref{fig:ray}, \ref{fig:ray-a0} and \ref{fig:r3m-ray} in \S
2 and 3. First, toroidal velocity of (either a massive or a massless)
particle seen by a local observer in LNRF of Kerr metric is given by
\begin{eqnarray}
v^{\hat{r}} &=& \frac{A^{1/2}}{\Delta} \frac{\dot{r}}{\dot{t}} \ ,
\label{eq:A1} \\
v^{\hat{\phi}} &=& \frac{A}{\Delta^{1/2} \Sigma}
\left(\frac{\dot{\phi}}{\dot{t}}-\omega \right) \ , \label{eq:A2}
\end{eqnarray}
where we restrict ourselves to equatorial trajectories ($\theta =
\pi/2$) in the disk plane. In this frame the photon emission angle
$\delta$ between the propagation direction and the radial direction
\citep[e.g.][p. 675]{MTW73} is defined as
\begin{eqnarray}
\cot \delta \equiv \frac{v^{\hat{r}}}{v^{\hat{\phi}}} =
\frac{\Sigma} {(A \Delta)^{1/2}} \frac{\dot{r}}{\dot{\phi}-\omega
\dot{t}} \ . \label{eq:A3}
\end{eqnarray}
On the other hand, radial component of the geodesic equation is given by
\begin{eqnarray}
\Sigma^2 \dot{r}^2 = R(r) \ , \label{eq:A4}
\end{eqnarray}
where $R(r) = (r^2+a^2-ab)^2-\Delta (b-a)^2$ for equatorial
trajectories and $b$ is photon's impact parameter. With the help of
equations~(\ref{eq:tdot}) and (\ref{eq:phidot}) one can express $b$
in terms of $\dot{r}$ and $\delta$ as
\begin{eqnarray}
b (\dot{r},\delta) &=& \dot{r} \left(\frac{r
\xi}{\Delta}\right)^{1/2} \tan \delta \ . \label{eq:A5}
\end{eqnarray}
Substituting equation~(\ref{eq:A5}) into equation~(\ref{eq:A4}) one
obtains an explicit expression for $\dot{r}$ in terms of the local angle
$\delta$ only as
\begin{eqnarray}
\dot{r}_{\pm} (\delta) &=& \frac{\xi^{1/2} \cos^2 \delta \left(\pm r
\Delta |\sec \delta| - 2 a M \Delta^{1/2} \tan \delta \right)}
{r^{1/2} \left[r^3(r-2M)+a^2(r^2-2M^2)+2a^2M^2 \cos(2\delta)\right]}
\ , \label{eq:A6}
\end{eqnarray}
which is equation~(\ref{eq:rdot}), and the sign in the numerator
depends on the direction of photon emission in the rest-frame of
fluid/source. Correspondingly, equation~(\ref{eq:A5}) is now
rewritten as
\begin{eqnarray}
b_\pm (\delta) &=& \frac{\xi \sin(2\delta) [\pm r \Delta |\sec
\delta| - 2 a M \Delta^{1/2} \tan \delta]}{2 \Delta^{1/2}
\left[r^3(r-2M)+a^2(r^2-2M^2)+2a^2M^2 \cos(2\delta)\right]} \ .
\label{eq:A7}
\end{eqnarray}

\end{document}